\documentclass[11pt]{article}

\usepackage{amsmath,amsfonts,amssymb,amsthm,bbm,stmaryrd}
\usepackage[dvips]{graphicx}
\usepackage{latexsym}
\usepackage{psfrag}
\usepackage[latin1]{inputenc}
\usepackage{hyperref}
\usepackage{color}
\usepackage{enumerate}
\numberwithin{equation}{section}
\usepackage{float}
\usepackage{cite}
\usepackage[T1]{fontenc}

\newcommand{\be}{\begin{equation}}
\newcommand{\ee}{\end{equation}}
\newcommand{\barray}{\begin{array}}
\newcommand{\earray}{\end{array}}
\newcommand{\bea}{\begin{eqnarray}}
\newcommand{\eea}{\end{eqnarray}}
\newcommand{\bs}{\begin{subequations}}
\newcommand{\es}{\end{subequations}}

\newcommand{\bit}{\begin{itemize}}
\newcommand{\eit}{\end{itemize}}
\newcommand{\bd}{\begin{description}}
\newcommand{\ed}{\end{description}}

\def\nn{\nonumber}

\def\la{\langle}
\def\ra{\rangle}
\newcommand{\bra}[1]{\la {#1}|}
\newcommand{\ket}[1]{|{#1}\ra}

\newcommand{\p}{\partial}

\newcommand{\f}{\frac}

\renewcommand{\d}{\delta}  \newcommand{\eps}{\epsilon}

\let\m=\mu    \let\n=\nu

\newcommand{\dbar}{d\hspace*{-0.08em}\bar{}\hspace*{0.1em}}

\newcommand{\scri}{{\cal I}}

\DeclareMathOperator{\sinc}{sinc}

\begin{document}

\title{\bf Black hole thermodynamic potentials for asymptotic observers}

\author{\Large{Antoine Rignon-Bret}
\smallskip \\ 
\small{\it{Aix Marseille Univ., Univ. de Toulon, CNRS, CPT, UMR 7332, 13288 Marseille, France}} }

\maketitle

\begin{abstract}
    The generalized second law states the total entropy of any closed system as the Universe cannot decrease if we include black hole entropy. From the point of view of an asymptotic observer, a black hole can be described at late time as an open system at fixed temperature which can radiate energy and entropy to infinity. I argue that for massless free quantum fields propagating on a black hole background, we can define a black hole dynamical free energy using observables defined at future null infinity which decreases on the successive cross sections of 
    $\scri^+$. The proof of this spontaneous evolution law is similar to Wall's derivation of the generalized second law and relies on  the monotonicity properties of the relative entropy. I discuss first the simpler case of the Schwarzschild background in which the gray body factor are neglected and show that in this case the free energy only depends on the Bondi mass, the Hawking temperature and the von Neumann entropy of the propagating quantum fields. Then I argue that taking into account the gray body factors adds a new term to the thermodynamic potential involving the number of particles detected at $\scri^+$ conjugated to a chemical potential. Finally, I discuss the case of the Kerr black hole for which an angular momentum piece needs to be added to the free energy.   
\end{abstract}

\tableofcontents
\section{Introduction}

The generalized second law has been introduced by Bekenstein \cite{bekenstein2020black} in order to save the ordinary second law of thermodynamics in the presence of black holes. Roughly speaking, it states that the total entropy of the Universe, which is a closed system, cannot decrease. It is the same statement as the ordinary second law of thermodynamics, except that now we can include the entropy of the black holes, which is proportional to the black hole horizon areas. Therefore, the sum of the black holes entropy and the entropy of the quantum field outside cannot decrease. Many authors have come with illuminating ideas to derive it from the first principles since Bekenstein (see \cite{Wall:2009wm} for a nice review), but the most complete proof of the generalized second law, as far as I know, is Wall's proof \cite{Wall:2010cj, Wall:2011hj}, which relies on the properties of the relative entropy, a fundamental quantity  in quantum thermodynamics and quantum information. The domain of application of the generalized second law is very broad, and it has been studied on future causal horizons \cite{Jacobson:2003wv}, in cosmology \cite{Bousso:2015eda}, at null infinity \cite{Kapec:2016aqd, Bousso:2016vlt, Bousso:2016wwu}, on general subregions \cite{Jensen:2023yxy, Ciambelli:2024fdgsl} and hides many clues about the nature of quantum gravity \cite{Wall:2010jtc}. In parallel, the nature of the generalized entropy has been subject to many investigations recently \cite{Witten:2021unn, Chandrasekaran:2022cip, Chandrasekaran:2022eqq, Jensen:2023yxy, Kudler-Flam:2023qfl, Faulkner:2024gst}.

\vspace{0.3 cm}

In the paper, we assume that the perturbed black hole is an isolated system that can be modeled in the framework of asymptotically flat spacetime. However, a realistic observer is never infinitely far from the black hole, and so we should expect some corrections to the formulas derived in the paper that depend on the dimensionless parameter $\frac{r_S}{r}$, $r_S$ being the black hole radius. If $r >> r_S$, then those corrections are negligible. In addition, in a more realistic spacetime, as an asymptotically de Sitter spacetime, there is also a cosmological horizon with an associated temperature. As long as we can neglect the presence of the cosmological horizon, the calculations presented in this paper are a very good approximation of the dynamics of the black hole as seen by a distant observer.

\vspace{0.3 cm}

\begin{figure}[t]
\begin{center}
\includegraphics[scale=0.9]{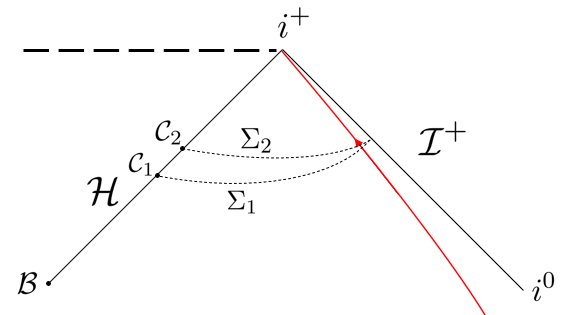}
\caption{An asymptotic observer (whose world line is depicted in red here) cannot make sense of the black hole area at retarded time $u$. Indeed, there is no canonical spacelike slice joining the asymptotic observer and the black hole event horizon. Indeed, the slices $\Sigma_1$ and $\Sigma_2$ have the same endpoint on $\scri^+$ while the cross sections $\mathcal{C}_1$ and $\mathcal{C}_2$ have, in general, different areas, so that the notion of the black hole area at retarded time $u$ for the asymptotic observer is meaningless: the dynamics of the horizon are not observable from future null infinity. }
\label{observerpenrosediag}
\end{center}
\end{figure}

We can draw the Penrose diagram depicting this observer far away from the black hole and located in a lab on planet Earth for instance, see Figure  \ref{observerpenrosediag}. 
We can assume that the asymptotic observer follows the world line spanned by the asymptotic vector $\p_u \lvert_r$, where $u$ is the retarded time of the usual Bondi coordinates in the conformal completed spacetime (where we added the null boundary $\scri$ to the spacetime manifold). We can notice that  there are an infinite number of spacelike slices $\Sigma$ containing the observer on $\scri^+$ at "time" $u$ and crossing the black hole horizon $\mathcal{H}$, and that there is no canonical way of choosing one over another.  Therefore, it is impossible for the asymptotic observer to associate an area $A$ to the black hole, since, in general, it will depend on the choice of cut of the horizon. One way of proceeding would be to measure the asymptotic mass and angular momentum of the black hole, and use the black hole equation of state, the Smarr formula \cite{Smarr:1972kt}
\be
    \mathcal{M} =  \frac{\kappa (\mathcal{M}, \mathcal{J})}{4 \pi} \mathcal{A}_\mathcal{H} + \Omega_\mathcal{H} \mathcal{J}   
    \label{smarr}
\ee
\footnote{This formula is true if there is no matter between the horizon and the observer. If there is, then we should take it into account in the Smarr formula} to deduce the horizon area from the asymptotic charges $\mathcal{M}$ and $\mathcal{J}$. Nevertheless, this formula is valid for stationary black holes. However, we are interested in the generalized second law, which is trivial as long as the black hole area is constant. If the black hole is perturbed by throwing some matter inside, then its area changes by an amount $\Delta \mathcal{A}$ between two cross sections $\mathcal{C}_1$ and $\mathcal{C}_2$ on the event horizon, and therefore, the notion of the black hole area as measured by an asymptotic observer becomes meaningless, since the spacelike slices $\Sigma_1$ and $\Sigma_2$ cross $\scri^+$ at the same cross section but represent different "times" on the black hole horizon, see Figure \ref{observerpenrosediag}. 

\vspace{0.3 cm}

Nevertheless, it is tempting to ask if there exists a way to check the second law for these asymptotic observers, despite the fact that they do not have access to the black hole area. By definition of what a black hole is, the asymptotic observers cannot have access to the local geometry of the event horizon, but they have access to the asymptotic geometric quantities. An idea that we can pursue is that at late time after the collapse, the black hole acts almost like a thermal reservoir with temperature $T_H$ with respect to the asymptotic observer if there were no incoming particles at past null infinity \cite{Hawking:1974rv,hawking1975particle, Wald:1975kc} \footnote{We say "almost" because of the gray body factors. We take into account the gray body factors in section \ref{graybodysection} but we can omit them for now.}. Therefore, in general, the black hole can emit electromagnetic radiation and gravitational waves which are registered and studied by the asymptotic observer in a thermal background at temperature $T_H$. \footnote{Of course, this is the black hole surrounding which emits radiation and particles, the black hole region itself causally connected from future null infinity.} From this perspective, going from the event horizon to future null infinity is equivalent to a change of thermodynamic variables and thus to a change of thermodynamic potential in order to study the spontaneous evolution. This shift is also motivated by the many similarities between the structures of horizons and null infinity \cite{Ashtekar:2024bpi, Ashtekar:2024mme} and the existence of monotonic fluxes of classical observables both at future null infinity and on the black hole event horizon \cite{Bondi:1960jsa, Hawking:1971tu, Rignon-Bret:2023fjq, Ciambelli:2023mir, Rignon-Bret:2024mef}.

\vspace{0.3 cm}

In Section \eqref{basictherm}, we  review basic thermodynamics of ordinary systems in order to motivate our change of variables and we make the connection with black hole physics.
In Section \ref{derivation} we prove in the simpler case, where the gray body factors are neglected, that if some massless and noninteracting quantum fields evolve in a Schwarzschild black hole background, then the following spontaneous evolution law holds
\be
    \Delta (\mathcal{M}_i - 8 \pi G T_H S_{\Sigma_i}(\rho)) \leq 0
    \label{inequalityfreeneergy}
\ee
 where $\mathcal{M}_{i}$ is the "geometric" Bondi mass \footnote{which is just the usual Bondi mass up to a factor $8 \pi G$.} computed at the cross section $\mathcal{C}_i$, and $S_{\Sigma_i}(\rho))$ is the von Neumann entropy of the quantum fields in the state $\rho$ evaluated on a spacelike slice $\Sigma_i$ intersecting null infinity at the arbitrary cross section $\mathcal{C}_i$. The derivation is very similar to Wall's proof of the generalized second law \cite{Wall:2010cj, Wall:2011hj}, but applied here on portions of $\scri^+$ instead of  $\mathcal{H}$. The idea behind this change of variables is essentially the same as in ordinary thermodynamics. For an isolated system, the study of entropy tells us about the spontaneous evolution of our physical system. However, if the system interacts with us by exchanging energy or particles, as the gravitational system does by emitting radiation having energy, entropy and angular momentum, then free energy is a more suitable choice as a thermodynamic potential compared to entropy, and is the natural candidate to tell us about the spontaneous evolution of the system. 

\vspace{0.3 cm}

In Section \ref{graybodysection}, we show that the influence of the gray body factors can be taken into account by adding another conjugate pair to the thermodynamic potential \eqref{inequalityfreeneergy} involving the average number of particles in each mode and their corresponding chemical potential, that we introduce here and which depends on the gray body factors. In this case, the inequality \eqref{inequalityfreeneergy} is  shifted into
 \be
    \Delta (\mathcal{M}_{i} - 8 \pi G T_H S_{\Sigma_i}(\rho) - \sum_{\omega, l m} \m_{\omega l m} \langle N_{\omega l m} \rangle_\rho) \leq 0.
    \label{inequalityfreeneergy2}
\ee
The appearance of a chemical potential can be understood by noticing that the number of particles propagating in the stationary background from the event horizon to future null infinity is indeed not conserved, since some of them will be reflected back and absorbed by the black hole because of the gravitational potential. From the point of view of an asymptotic observer, these particles are destroyed as in an ordinary chemical reaction, and this is why a chemical potential appears naturally in the spontaneous evolution law.  Finally, in Section \ref{kerrsection} we treat the case of the Kerr black hole, in which \eqref{inequalityfreeneergy2} is modified into 
 \be
    \Delta (\mathcal{M}_{i} - \Omega_H \mathcal{J}_{i} - 8 \pi G T_H S_{\Sigma_i}(\rho) - \sum_{\omega, l m} \m_{\omega l m} \langle N_{\omega l m} \rangle_\rho) \leq 0
    \label{inequalityfreeneergy3}
\ee
and discuss the classical limit $\hbar \rightarrow 0$ and the vanishing temperature limit $\kappa \rightarrow 0$.

\vspace{0.3 cm}

However, we will not discuss the potential divergences that we could encounter in the derivation and assume that they do not represent major obstacles. First, it should be noticed that for quantum field theories, the entanglement entropy diverges and so the quantity 
\be
    \mathcal{F}_i = \mathcal{M}_{i} - 8 \pi G T_H S_{\Sigma_i}(\rho)
    \label{f function}
\ee
is not well defined \textit{a priori} \footnote{However we also need to renormalize the Newton constant, such that the expression \eqref{f function} could actually be well defined, as it is the case for the generalized entropy \cite{susskind1994black}}. This problem is often treated by renormalizing the von Neumann entropy by subtracting the entropy of the vacuum (here it would be the asymptotic "out" vacuum) in order to get finite results. However, for differences as von Neumann entropies as it is the case in the spontaneous evolution laws \eqref{inequalityfreeneergy}, \eqref{inequalityfreeneergy2}, and  \eqref{inequalityfreeneergy3}, it is not a problem \textit{a priori}. Furthermore, Gibbs states, that we will use as thermal states of reference, cannot be well defined in quantum field theories because of the infinite amount of degrees of freedom coming into play \footnote{Notions as the trace are ill-defined for quantum field theories.}. As we write our thermal states as Gibbs states in the following, the arguments given here can only be formal. However, we can consider that we can regularize our quantum field theory, such that there is only a finite (but large) number of degrees of freedom remaining, and assume that the results also hold when we take the limit in the case of an infinite amount of degrees of freedom. 

\vspace{0.3 cm}

This work is not the first attempt to give a version of the generalized second law at $\scri^+$, as it has been shown that a generalized second law holds for a renormalized area \cite{Kapec:2016aqd} and that quantum bounds have already been worked out \cite{Bousso:2016vlt, Bousso:2016wwu} using the quantum null energy conditions \cite{Bousso:2015wca}. However, this work differs from the previous ones by focusing on the asymptotic quantum fields in a black hole background spacetime. Indeed, the results of this paper are based on the states of the quantum fields at late time at $\scri^+$ after a black hole collapse \cite{hawking1975particle, Wald:1975kc}, which is not the case for any of the works mentioned above.

\vspace{0.3 cm}

In this paper we will work with units $c = k_B = 1$.

\section{Basics of thermodynamics and black holes}
\label{basictherm}
Let us consider now a small system $S$ with total energy $\mathcal{E}$ and volume $\mathcal{V}$. It can exchange volume and energy with a large reservoir $\mathcal{R}$ with temperature $\mathcal{T}_\mathcal{R}$ and pressure $\mathcal{P}_\mathcal{R}$. The first law of thermodynamics tells us that the energy variation $\mathcal{E}$ of the small system $S$ is given by
\be
    d \mathcal{E} = \dbar \mathcal{W} + \dbar \mathcal{Q} = - \mathcal{P}_\mathcal{R} d \mathcal{V} + \dbar \mathcal{Q}
    \label{firstlaw}
\ee
where $\dbar \mathcal{W}$ and $\dbar \mathcal{Q}$ are, respectively, the work and heat injected into the small system $S$, and $\mathcal{V}$ is the volume of $S$. Now we can combine the first law \eqref{firstlaw} with the second law telling us that there exists a function of the state of the system $\mathcal{S}$ called the entropy which variation is given by
\be
    d \mathcal{S} = \frac{\dbar \mathcal{Q}}{\mathcal{T}_\mathcal{R}} + \dbar \mathcal{S}_c
    \label{secondlaw}
\ee
where $\mathcal{Q}$ is the heat current flowing into the system $S$ and $\dbar \mathcal{S}_c$ is the entropy creation term with 

\be \label{spontevol}
\dbar \mathcal{S}_c \geq 0.
\ee
The condition \eqref{spontevol} tells us about the spontaneous evolution of the system $S$. Indeed, the quantity $\dbar \mathcal{S}_c$ vanishes if the system is in internal and external equilibrium but is strictly positive otherwise, i.e if there is an internal gradient of temperature or pressure for instance. Therefore, it tells us about the spontaneous evolution of our physical system. Hence, for an isolated system $\mathcal{Q} = 0$ and we deduce that $\mathcal{S}$ can only increase \footnote{It is the case for the Joule Gay-Lussac expansion for instance, since the entropy increases while there is no heat exchange with the exterior.}. Implementing \eqref{secondlaw} into \eqref{firstlaw}, we get
\be
    \mathcal{T}_\mathcal{R} d \mathcal{S} - d \mathcal{E} - \mathcal{P}_\mathcal{R} d \mathcal{V} =  \mathcal{T}_\mathcal{R} \dbar \mathcal{S}_c \geq 0.
    \label{firstinequlaitytherm}
\ee
Therefore, in the case where the system $S$ does not exchange neither energy nor volume with its environment, we have $d \mathcal{E} = d \mathcal{V} = 0$ and so $d \mathcal{S} \geq 0$. Replicas of this isolated system $S$ form the microcanonical ensemble. In this case, the entropy is a suitable thermodynamic potential because it allows us to follow the system evolution : it only increases with time, and thermodynamic equilibrium is reached for a maximum amount of entropy.

Now, let us consider that the system $S$ can exchange energy with its environment. We can rearrange \eqref{firstinequlaitytherm} and write 
\be \label{potthermo}
    d (\mathcal{E} - \mathcal{T}_\mathcal{R} \mathcal{S}) + \mathcal{S} d \mathcal{T}_\mathcal{R} + \mathcal{P}_\mathcal{R} d \mathcal{V} = - \mathcal{T}_\mathcal{R}\dbar S_c \leq 0
\ee
and so, if we work at fixed volume and fixed temperature of the reservoir $ d \mathcal{T}_\mathcal{R} = d \mathcal{V} = 0$, then the quantity $\mathcal{F} = \mathcal{E} - \mathcal{T}_\mathcal{R} \mathcal{S}$ decreases over time. The quantity $\mathcal{F}$ is not exactly the free energy of the system since it can be defined even if the system is not in thermal equilibrium, (and thus does not have a well-defined temperature), as long as the reservoir is in thermal equilibrium and has a well-defined temperature. It becomes the free energy if $\mathcal{T} = \mathcal{T}_\mathcal{R}$. However, in what follows we will be abusing the terminology a little by referring to these two quantities as free energy.  Replicas of this system $S$ form the canonical ensemble, and $\mathcal{F}$ is a suitable thermodynamic potential in order to study the spontaneous evolution of the system at fixed volume and fixed external temperature, as it decreases monotonically in time. Notice that if we do not fix the number of particles of each species in the system, by allowing chemical reactions for instance, then we have to add an additional term $\sum_i \m_{\mathcal{R}, i} \mathcal{N}_i$ to \eqref{firstlaw} and in this case \eqref{potthermo} becomes
\be \label{potthermo1}
    d (\mathcal{E} - \mathcal{T}_\mathcal{R} \mathcal{S} - \sum_i \m_{\mathcal{R}, i} \mathcal{N}_i) + \mathcal{S} d \mathcal{T}_\mathcal{R} + \mathcal{P}_\mathcal{R} d \mathcal{V} + \sum_i  \mathcal{N}_i d \m_{\mathcal{R}, i} = - \mathcal{T}_\mathcal{R}\dbar S_c \leq 0
\ee
where $\mathcal{N}_i$ is the number of particles of species $i$ in the system. Therefore, if we add the condition $d  \m_{\mathcal{R}, i} = 0$ in addition to $ d \mathcal{T}_\mathcal{R} = d \mathcal{V} = 0$, then we find that the quantity $\mathcal{F} = \mathcal{E} - \mathcal{T}_\mathcal{R} \mathcal{S} - \sum_i \m_{\mathcal{R}, i} \mathcal{N}_i$ decreases over time. It is sometimes referred as $\mathcal{J}$ and called the grand potential, but we will continue to name it free energy and denote it $\mathcal{F}$ in the following in order to avoid confusion, even if it is indeed a different thermodynamic potential. Replicas of the system $S$ form now the grand canonical ensemble, and $\mathcal{J}$ is the suitable thermodynamic potential in order to study its spontaneous evolution. 

\vspace{0.3 cm}

\begin{figure}[t]
\begin{center}
\includegraphics[scale=0.9]{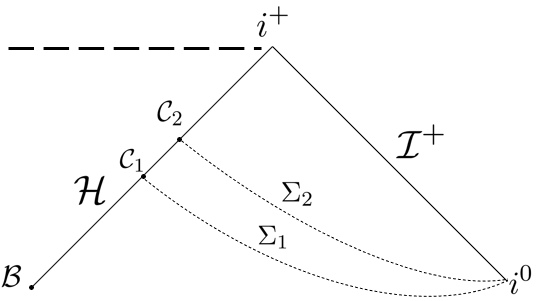}
\caption{Here two spacelike $\Sigma_1$ and $\Sigma_2$ intersect the black hole horizon at the corners $\mathcal{C}_1$ and $\mathcal{C}_2$ respectively, and spatial infinity $i^0$ (the generalized second law would hold if the slices $\Sigma_i$ intersect any corner $\mathcal{C}$ of $\scri^+$ as long as it is the same corner $\mathcal{C}$ for all the spacelike slices $\Sigma_i$). The cut $\mathcal{B}$ can be thought as the bifurcation surface, but not necessarily. }
\label{entropytp}
\end{center}
\end{figure}

If black holes come into the game, we have at our disposal the generalized second law, which states that the total entropy of the Universe (the black hole entropy plus the entropy of the matter outside) increases over time. Here, time has a very specific meaning; it is meant to be a family of disjoint spacelike slices $\Sigma_v$ labeled by one parameter $v$ \footnote{We can identify this parameter with an affine coordinate labeling the null geodesics on the black hole event horizon.}intersecting the black hole event horizon at the corner $\mathcal{C}_v$ and spacelike infinity $i^0$. They are ordered in time such as if $v_2 > v_1$, $\Sigma_{v_2}$ is in the future of $\Sigma_{v_1}$, see Figure \ref{entropytp}. The generalized second law states that

\be
\frac{\mathcal{A}_{\mathcal{C}_{v_2}}}{4 G \hbar} + S^{vN}_{\Sigma_{v_2}} \geq \frac{\mathcal{A}_{\mathcal{C}_{v_1}}}{4 G \hbar} + S^{vN}_{\Sigma_{v_1}}
\ee
where $S_{\Sigma_{v_i}} = - Tr(\rho \ln{\rho}) \lvert_{\Sigma_{v_i}}$ is the von Neumann entropy of the quantum fields outside the black hole. Since all the Cauchy slice $\Sigma_{i_1}$ have spacelike infinity as an end point \footnote{It is important to realize that the generalized second law would still be correct if the spacelike slices had the same end point $\mathcal{C}$ on $\scri^+$, but \textit{not} if the endpoint is a codimension two surface in the bulk.}, the system is closed, in the sense that there exists no physical processes such that energy and angular momentum which can be radiated away at spacelike infinity even if there can be, of course, radiation escaping at future null infinity (see Figure \ref{entropytp}).
The successive spacelike slices keep track of all the information outside the black hole event horizon as long as it does not fall into the black hole. Indeed, the generalized second law is a statement about the entropy of the Universe (containing black holes, otherwise it is just the ordinary second law) as a whole. Of course, as long as the spacelike slices share the same end point at (spacelike or future null) infinity, the mass and angular momentum, which can be computed on this asymptotic corner as Noether charges \cite{wald1993black, iyer1994some, iyer1995comparison, wald2000general}, remain constant, and so the black hole is analogous to a system with fixed energy and volume that we studied previously.

\begin{figure}[t]
\begin{center}
\includegraphics[scale=0.9]{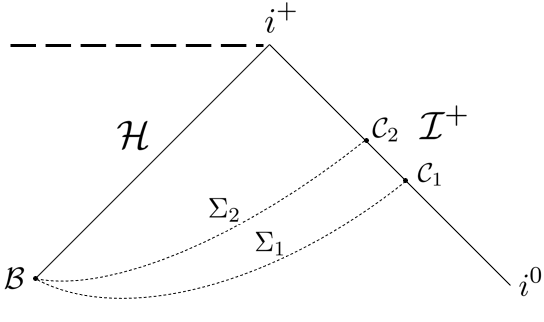}
\caption{Here two spacelike slices $\Sigma_1$ and $\Sigma_2$ intersect null infinity $\scri^+$ at the corners $\mathcal{C}_1$ and $\mathcal{C}_2$ respectively, and the black hole at the corner $\mathcal{B}$ (not necessarily the bifurcation surface, which does not exist for a real scenario collapse). The end point $\mathcal{B}$ can be chosen in a regime such that the event horizon is almost a Killing horizon, but not necessarily.}
\label{freeenergytp}
\end{center}
\end{figure}

However, from the point of view of an asymptotic observer (or a family of asymptotic observers), the black hole is not an isolated object. As we explained in the Introduction, the black hole surrounding can emit electromagnetic or gravitational waves. 
These emissions include  Hawking radiation. One feature that far away observers can (in principle) detect about the black hole is its almost (it is exactly thermal only if one neglects the gray body factors) thermal spectrum. Therefore, for asymptotic observers, at late retarded time $u$, the black hole behaves like a reservoir at temperature $\mathcal{T}_{\mathcal{R}} = T_H = \frac{\hbar \kappa}{2 \pi}$ \cite{hawking1975particle, Wald:1975kc}. From this point of view, and on time scales for which the back reactions can be neglected and we can truly consider the black hole as a reservoir at temperature $T_H$, (we indeed have that $d T_H \sim \frac{d M}{M^2} << T_H$ for large black holes) it seems more reasonable for the asymptotic observer to treat the black hole background spacetime and field perturbations on top of it as an open system at fixed temperature, which can exchange mass and angular momentum with the outside (i.e which can emit energy and angular momentum to null infinity), rather than as a closed system. Anyways, it is the way in which far away observers working in their lab should perceive the black hole, as a hot environment emitting radiation and exchanging energy and entropy. Therefore, the thermodynamic potential that is relevant for the asymptotic observer is not the generalized entropy but a quantity similar to the free energy $\mathcal{F}$. Thus, contrary to what was done for the generalized second law, we now consider a region of future null infinity bounded by two cross sections $\mathcal{C}_1$ and $\mathcal{C}_2$ (see Figure \ref{freeenergytp}) and are interested in the energy and entropy flowing through this region and not on a portion of the event horizon $\mathcal{H}$. However, we should notice the many similarities between the figures \ref{entropytp} and \ref{freeenergytp}. In the end, the main difference between the two pictures can be summarized by the choice of boundary conditions, either we are interested in the dynamics at null infinity or on the horizon in the black hole background spacetime. 

\vspace{0.3 cm}

In the remainder of the paper, except for the last section \eqref{kerrsection} where we treat Kerr, we focus on the Schwarzschild black hole, with no angular momentum. Usually, the quantity $\mathcal{F}$ decreases for constant temperature and constant volume. However, we allow the quantum fields to carry angular momentum to infinity, but impose that $\Omega_H = 0$, which thus would be similar to the condition $\mathcal{P}_{\mathcal{R}} = 0$ in the classical thermodynamic system and not to the usual fixed volume condition $d \mathcal{V} = 0$. Therefore, the quantum fields can carry angular momentum, but they still propagate on a Schwarzschild background. In section \eqref{kerrsection}, they  propagate in a Kerr background.

\section{Derivation of the spontaneous evolution law}
\label{derivation}
\begin{figure}[t]
\begin{center}
\includegraphics[scale=0.9]{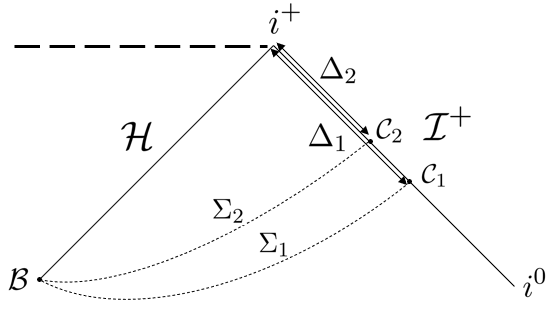}
\caption{Here two spacelike slices $\Sigma_1$ and $\Sigma_2$ intersect null infinity $\scri^+$ at the corners $\mathcal{C}_1$ and $\mathcal{C}_2$ respectively, and the black hole at the corner $\mathcal{B}$ (not necessarily the bifurcation surface, which does not exist for a real scenario collapse). $\Delta_i$ is the portion of $\scri^+$ between the asymptotic corner $\scri^+_+$ which is the future of future null infinity and the corner $\mathcal{C}_i$. }
\label{Figurelightcone}
\end{center}
\end{figure}

Let us consider an observer in a laboratory very far away from the black hole and at very late time after the collapse. Throughout this section, we assume that the black hole in question is extremely large, so that the gravitational potential barrier is small outside the black hole and we can neglect the gray body factors in the analysis\footnote{In that case we could argue that the black hole temperature $T_H$ is also infinitesimally small. However, we face a textbook case here and we allow ourselves to modify the Planck constant so that $\frac{\hbar}{M}$ remains of order $1$ even though $M$ is arbitrarily large. In this setting, the temperature can be finite even though the black hole is very large.}. However, taking into account the presence of gray body factors is not a big deal and we will do it in Section \ref{graybodysection}. Nevertheless, in order to understand better the key ideas, it seems to be better to ignore them in the first instance.

\vspace{0.3 cm}

In this setting, the black hole background acts as a thermal reservoir for our observer with temperature $T_H$. Let us consider two spacelike slices $\Sigma_1$ and $\Sigma_2$ intersecting the event horizon at the cross section $\mathcal{B}$ and two cross sections $\mathcal{C}_1$ and $\mathcal{C}_2$ at null infinity (see Figure \ref{Figurelightcone}). Let $\Delta_{1}$ ($\Delta_{2}$) be the portion of $\mathcal{I}^{+}$ going from $\scri^+_+$ (at $u \rightarrow + \infty$ in adapted Bondi coordinates in a neighborhood of $\mathcal{I}^+$) to $\mathcal{C}_1$ ($\mathcal{C}_2$). Furthermore, let us define $\bar{\Delta}_i$ such that $\Delta_i \cup \bar{\Delta}_i = \scri^+$ and $\Delta_i \cap \bar{\Delta}_i = \emptyset $. In order to derive the spontaneous evolution law \eqref{inequalityfreeneergy}, we use the notion of relative entropy between two quantum states $\rho$ and $\sigma$. The relative entropy is defined as
\be \label{relativeentropy0}
    S(\rho \lvert \lvert \sigma) = Tr(\rho \ln{\rho}) - Tr(\rho \ln{\sigma})
\ee
and tells us how far apart the two states $\rho$ and $\sigma$ are. The relative entropy \eqref{relativeentropy0} is non-negative and vanishes if and only if $\rho = \sigma$. In order to prove the inequality \eqref{inequalityfreeneergy}, we use a well-known property of monotonicity  of the relative entropy which was the key point in Wall's derivation of the generalized second law \cite{Wall:2010cj,Wall:2011hj}. This property states that if we consider two achronal regions $A$ and $B$  such that $A \subset B$, then \cite{lindblad1975completely} 
\be
    S(\rho \lvert \lvert \sigma) \lvert_A \leq S(\rho \lvert \lvert \sigma) \lvert_B
    \label{relativeentropy}
\ee
Intuitively, it means that two states $\rho$ and $\sigma$ are less distinguishable if we restrict ourselves to the observables defined in the region $A$ instead of considering all the observables available in the region $B$ since the observables in region $A$ are all included in region $B$. We will apply the inequality \eqref{relativeentropy} to the portions of $\mathcal{I}^+$ $A = \Delta_2$ and $B = \Delta_1$ in Figure \ref{Figurelightcone}.

\vspace{0.3 cm}

We will compare the states corresponding to an arbitrary state of our quantum field $\rho$ and to a reference state $\sigma$, at late retarded time $u$ on $\scri^+$. The reference state $\sigma$ is obtained through the unitary evolution in a black hole collapse background of the vacuum of $\scri^{-}$ $\sigma_0 = \ket{0}_{-} \bra{0}_-$  and restricting it to late time on $\scri^+$, while the quantum state $\rho$ is obtained from the unitary evolution on the black hole collapse background of an arbitrary state $\rho_0$ on $\scri^{-}$ restricted to the same asymptotic late region of $\scri^+$. It has been shown in the seminal papers \cite{hawking1975particle, Wald:1975kc}  that at late time on $\scri^+$ , if we neglect the influence of the gravitational potential and the gray body factors, on which we come back to in Section \ref{graybodysection}, then the quantum field is in the formal Gibbs state
\be
    \sigma \underset{u \rightarrow + \infty}{\sim} \mathbb{P}_u  \frac{e^{-\beta_H \hat{H}}}{Tr(e^{-\beta_H \hat{H}})}
    \label{gibbsstate}
\ee
if we started in the \textit{in}-vacuum $\sigma_0$ at $\scri^-$. In \eqref{gibbsstate} $\beta_H = \frac{1}{T_H}$ is the inverse of the black hole temperature, $\mathbb{P}_u$ is a (trace preserving) projector on the states of $\mathcal{H}_{\scri^+}$ peaked at time $u$ and $\hat{H}$ is the (eventually renormalized) Hamiltonian of our quantum field, equal to 
\be \label{hamiltonian}
        \hat{H} = \int_{\scri^+} \hat{T}_{\m \n} n^\m n^\n \eps_{\scri^+}
\ee
where $n^\m = \p_u^\m$ is an affinely parametrized normal on $\scri^+$ and $\eps_{\scri^+}$ is the induced volume form on $\scri^+$ \footnote{Here $\eps_{\scri^+}$ is the unphysical induced volume form which is finite at $\scri^+$ and written in terms of the unphysical metric $g_{\m \n} = r^{-2} \tilde{g}_{\m \n}$ where $\tilde{g}_{\m \n}$ is the physical metric. Of course, the stress energy tensor $T_{\m \n}$ is therefore related to the physical stress energy tensor $\tilde{T}_{\m \n}$ via $T_{\m \n} = r^2 \tilde{T}_{\m \n} = O(1)$.}.

\vspace{0.3 cm}

\begin{figure}[t]
\begin{center}
\includegraphics[scale=0.4]{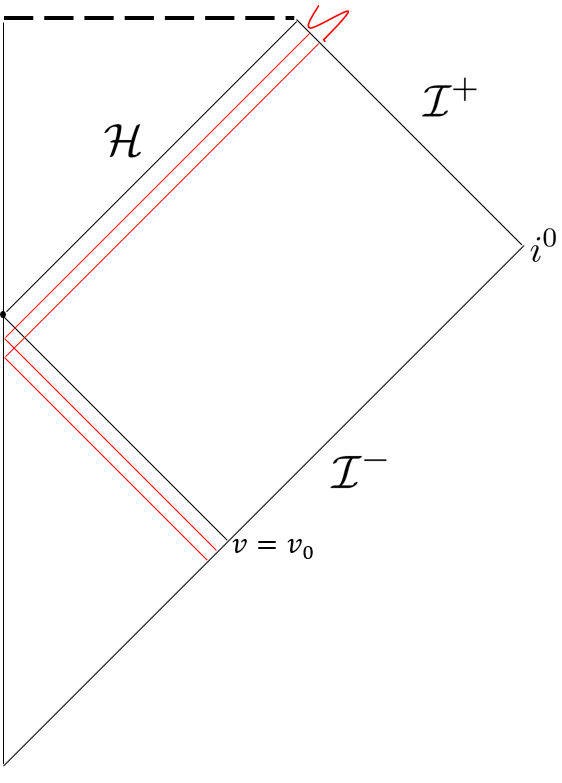}
\caption{The trajectory of a wave packet in a black hole spacetime formed through a collapse. If we take a wave packet with compact support at late retarded time picked around some frequency $\omega_0$, we can make it propagate backward in time until it reaches $\scri^-$. The affine coordinates $v_1$ and $v_2$ at $\scri^{-}$ are mapped to $u(v_1)$ and $u(v_2)$ following the red null geodesics. The late time assumption is essential because it is only in this region that $u(v_2) - u(v_1) = - 4 M \ln{\frac{v_0 - v_2}{v_0 - v_1}}$. This logarithmic relation is a the heart of the Hawking effect. It implies that there is an arbitrary large blue shift when the wave packet propagates backward in time from late times on $\scri^+$ to $\scri^-$. We should not forget that we disregarded the gravitational potential here such that there is no back scattering and all the light rays emerging from $\scri^+$ go through the collapsing body before reaching $\scri^-$ at $v < v_0$.
}
\label{wavepackettrace}
\end{center}
\end{figure}

The famous result \eqref{gibbsstate} is obtained by computing the Bogoliubov coefficients relating a natural decomposition of the \textit{in}-Hilbert space $\scri^{-}$ to a natural decomposition in the \textit{out}-Hilbert space, at late time on $\scri^+$.  The Bogoliubov coefficients can be computed as in Hawking \cite{hawking1975particle} and Wald \cite{Wald:1975kc}  original derivations by considering a wave packet centered in a late time region on $\scri^+$ where the background spacetime is assumed to be stationary, propagating backwards in time.
In general, one part of the wave packet is reflected by the Schwarzschild potential towards $\scri^-$ and there is no mixing of frequencies there, since we can approximate that the wave packet propagates on the stationary Schwarzschild background and so the Killing frequency $\omega$ is a conserved physical property (again, we disregard this contribution for now since we assume that the gray body factors vanish in this section). However, one part of the wave packet is transmitted through the potential barrier and comes very close to the horizon, where it is blue shifted, enters the collapsing body and reaches past null infinity with a different Fourier decomposition with respect to the time $v$ at past null infinity, see Figure \ref{wavepackettrace}. This shift in frequencies lies at the heart of the Hawking effect. From these relations, we can build a map relating the asymptotic past states to the states at late time on $\scri^+$. If the state on $\scri^-$ is the vacuum, then the state on $\scri^+$ is a mixed state given by \eqref{gibbsstate}. 

\subsection*{Hilbert space and orthonormal basis}
\label{Hssection}

In order to define properly the Gibbs state on a late time portion $\Delta_i$ of $\scri^+$, we have to write eigenstates of the Hamiltonian operator \eqref{hamiltonian} in terms of square integrable functions with compact support on $\Delta_i$. Indeed, since the Hamiltonian  \eqref{hamiltonian} involves an integral of the stress energy tensor of our free field on $\scri^+$, the eigenstates are not localized on $\Delta_i$ in general.  For simplicity, we assume here that $\Delta_i = ]0, +\infty[ \times S^2$, which can always be achieved by shifting the location of $\mathcal{C}_i$ using a supertranslation \footnote{We must be careful that the choice $u = 0$ is just a shift of origin and that we are still at "late time" even at $u = 0$, and by this we mean that $u = 0$ takes place a long time after the collapse, such that the background spacetime is stationary at $u \geq 0$ in the vicinity of $\scri^+$.}. The Hilbert space $\mathcal{H}_{\Delta_i}$ can be constructed out of the $n$-particles state

\be
    \mathcal{H}_{\Delta_i} = L^2 (\Delta_i) \oplus \big( L^2 (\Delta_i) \otimes L^2 (\Delta_i) \big) \oplus \cdots
\ee
where symmetrization is kept implicit and we can eventually add additional degrees of freedom such as the spin or helicity of the field considered. An orthonormal basis of $\mathcal{H}_{\Delta_i}$ is made of the states $f_{\omega_0, l, m}(u)$ where \footnote{Usually, there is a factor $r^{-1}$ since the modes must have finite energy at $\scri^+$ and must be normalized but we decided to include it instead in the definition of the volume form at $\scri^+$ $\eps_{\scri^+}$ which is the \textit{unphysical} volume form at $\scri^+$ written in terms of the unphysical metric $g_{\m \n} = r^{-2} \hat{g}_{\m \n}$ where $\hat{g}_{\m \n}$ is the physical metric. Accordingly, we will take for consistency $T_{\m \n} = r^2 \tilde{T}_{\m \n}$ where $\tilde{T}_{\m \n}$ is the physical stress energy tensor.}  

\be \label{normalizedmodes}
f_{\omega_0, l, m}(u) =  \left\{
    \begin{array}{ll}
         \sqrt{2 \eps} Y_{lm} e^{i (\omega_0 + i \eps) u}  \qquad & \text{if $u > 0$}\\
        0 \qquad & \text{otherwise}
    \end{array}
\right.
\ee
where the functions $Y_{lm}(\theta, \phi)$ are the normalized spherical harmonics with $l \geq 0$  and$-l \leq m \leq l$ and where $\eps > 0$ and assumed to be very small for all  purposes \footnote{Here we have that $f_{\omega_0, l, m}(u)$ does not vanish in $u = 0$ so we may worry about the discontinuity here. However, we can always regularize the wave packets  such that it decreases $f_{\omega_0, l, m}(u)$ smoothly for $u \in ]0, \eps[$ and vanishes at $u = 0$ where $\eps$ is arbitrarily small without changing any of the relevant properties for the present discussion.} \footnote{In Hawking \cite{hawking1975particle} and Wald \cite{Wald:1975kc}, the basis on which the wave packets are decomposed is the set
\be \label{Hwwavepacket}
        f_{lm}(j, n) = \frac{1}{\sqrt{\eps}} \int_{j \eps}^{(j + 1) \eps} e^{-\frac{i 2 \pi n \omega}{\eps}} b_{\omega l m} d\omega
\ee
where $j$ is a positive integer introduced to label the frequency spectrum and $n$ is an integer introduced to label the retarded time on $\scri^+$. Indeed, the wave packets \eqref{Hwwavepacket} are localized in frequency around $\omega \approx j \eps$ and are centered around $u = \frac{2 \pi n}{\eps}$
\be
        f_{lm}(j, n) = \sqrt{\alpha \eps} e^{i(\eps u - 2 \pi n)(j + \f12)}\sinc{\f12 (\eps u - 2 \pi n)} Y_{lm}
\ee
Of course, the wave packets of interest are the ones forming a basis at late time on $\scri^+$, therefore for $n \rightarrow + \infty$. However, if these wave packets are centered around $u = \frac{2 \pi n}{\eps}$, they are not localized on the region $\Delta_i$, which motivates us to choose the orthonormal basis \eqref{normalizedmodes} instead of \eqref{Hwwavepacket}. 
}. 
The orthonormality property of the family \eqref{normalizedmodes} is obtained in the limit $\eps \rightarrow 0$ . We can interpret $\Delta u = \frac{1}{\eps}$ as the characteristic length of the wave packet, or $\eps$ the frequency range for which our detector is able to distinguish two different frequencies. In addition, the modes $b_{\omega l m} (u) = \sqrt{\alpha} Y_{l m} e^{i \omega u}$ are eigenstates of the Hamiltonian \eqref{hamiltonian} with eigenvalue $\hbar \omega$ for any arbitrary constant $\alpha$. However, they do not belong the Hilbert space $\mathcal{H}_{\scri^+}$ since they are not normalized. 
Hence, in order to have the spectral decomposition of the normalized modes \eqref{normalizedmodes}, we compute
\be \label{decomposition}
        \hat{f}_{\omega_0, l l', m m'} (\omega) = \int_{\scri^+} f_{\omega_0, l, m} b^\ast_{\omega, l', m'} \eps_{\scri^+} =  \d_{l l'} \d_{m m'}  \frac{ i \sqrt{2 \alpha \eps}}{\omega - \omega_0 + i \eps} 
\ee
and so 
\be
    \lvert \hat{f}_{\omega_0, l l', m m'} \rvert^2 (\omega) d \omega = \frac{2 \alpha}{\big( \frac{\omega- \omega_0}{\eps} \big)^2 + 1} \frac{1}{\eps} d \omega 
\ee
 Therefore, we have $ \lvert \hat{f}_{\omega_0, l l', m m'}  \rvert^2 (\frac{\omega - \omega_0}{\eps}) = 0$ if $\lvert \omega - \omega_0 \rvert >> \eps$  and $\lvert \hat{f}_{\omega_0, l l', m m'}  \rvert^2 (\frac{\omega - \omega_0}{\eps}) = 2 \alpha$ if $\lvert \omega - \omega_0 \rvert << \eps$ which confirms the interpretation of the parameter $\eps$ as the frequency range in which we cannot distinguish two different frequencies. Perfect detectors have $\eps \rightarrow 0$ because they can distinguish all the frequencies. In this limit,
\be
        \lvert \hat{f}_{\omega_0, l l', m m'} (\omega)\rvert^2 = 2 \pi \alpha \d (\omega - \omega_0) \d_{ll'} \d_{mm'} 
\ee
and therefore we can conclude that the orthonormal modes \eqref{normalizedmodes} on $\Delta_i$ have a well-defined frequency and energy, and can be interpreted as (normalized) one particle states localized in the region $\Delta_i$ of $\scri^+$. 

\vspace{0.3 cm}

Therefore, from the one particle state we can construct the normalized $n$ particle state $\ket{N_{\omega_0 l m}}$ (for arbitrary large $\Delta u = \frac{1}{\eps}$) 

\be \label{nparticlestate}
    \ket{N_{\omega_0 l m}} = \frac{1}{\sqrt{N_{\omega_0 l m} !}} \big( \sum_{l' m'} \int_{- \infty}^{+ \infty} \frac{1}{2 \pi \sqrt{\alpha}} \hat{f}_{\omega_0, l l', m m'} (\omega) \hat{a}^\dag _{\omega, l', m'} d\omega \big)^{N_{\omega_0, l, m}} \ket{0}_+
\ee
where $\ket{0}_+$ is the \textit{out}-vacuum state on $\scri^+$ and $a_{\omega, l, m}$ and $a^\dag_{\omega, l, m}$ are the ladder operators such that the Hamiltonian of our free field on $\scri^+$ \eqref{hamiltonian} can be written as 
\be \label{hamiltonian2}
    \hat{H} = \sum_{\omega l m} \hbar \omega a_{\omega, l, m}^\dag a_{\omega, l, m}
\ee
where the integral over the frequencies is here depicted as a sum in order to lighten the notation. Then, we can write the restriction of the Hamiltonian \eqref{hamiltonian2} to $\mathcal{H}_{\Delta_i}$ by using the normalized states \eqref{nparticlestate} with compact support on $\Delta_i$, so that 
\be \label{restricthamil}
    \hat{H} \lvert_{\Delta_i} = \sum_{\{ N_{\omega l m} \}} \big( \sum_{\omega l m}N_{\omega l m} \hbar \omega \big) \ket{\{ N_{\omega l m} \} } \bra{ \{ N_{\omega l m} \} }
\ee
and the degrees of freedom of $\bar{\Delta}_i$ are traced out. From \eqref{restricthamil} we deduce $\sigma \lvert_{\Delta_i}$ which can be formally written as
\be \label{gibbsatedeltai}
    \sigma \lvert_{\Delta_i} = \frac{e^{- \beta_H \hat{H} \lvert_{\Delta_i}}}{Tr(e^{- \beta_H \hat{H} \lvert_{\Delta_i}})}
\ee
but for our purpose we only need $\ln{\sigma \vert_{\Delta i}} = - \beta_H \hat{H} \vert_{\Delta_i} + \ln{Z}$ since it is what appears in the relative entropy \eqref{relativeentropy0}. We should also notice that the normalization factor in \eqref{gibbsatedeltai}, equal to the logarithm of the partition function
\be
    \ln{Z} = Tr(e^{-\beta_H \hat{H}_{\Delta_i}})
\ee
will not be important in our discussion since we will be interested in \eqref{relativeentropy} and therefore in differences of relative entropy. 

\vspace{0.3 cm}

As we have already explained, tracing back in time the wave packets \eqref{nparticlestate} allows us to find a map between the states in $\mathcal{H}_{\scri^-}$ and the states in $\mathcal{H}_{\Delta_i}$ which are here states localized at late time on $\scri^+$ with a defined frequency in the limit $\eps \rightarrow 0$, that we have considered here. If we assume that the initial state of the quantum field at $\scri^-$ is $\sigma_0 = \ket{0}_{-} \bra{0}_{-}$, then the image of this map in $\mathcal{H}_{\Delta_i}$ is the state \eqref{gibbsatedeltai}. This mixed state indicates the probability of detecting $N_{\omega l m}$ particles  with quantum numbers $(\omega,l,m)$ if an asymptotic observer turns on a detector of resolution $\eps$ at $u = 0$ and leaves it running during the time interval $\Delta u = \frac{1}{\eps}$, as long as we choose $\eps$ to be very small and choose the origin $u = 0$ to be at very late time after the collapse, such that the background black hole spacetime has already settled down to a stationary state. In particular, any observable on $\Delta_i$ can be computed using \eqref{gibbsatedeltai} if we assume that the initial state at $\scri^-$ is $\sigma_0 = \ket{0}_{-} \bra{0}_{-}$.    

\subsection*{Supertranslation symmetry}
\label{stsection}

The symmetry by arbitrary supertranslations allows us to label any cross section $\mathcal{C}_i$ of $\scri^+$ as $u = 0$. Indeed, as for any pair of cross sections $\mathcal{C}_1$ and $\mathcal{C}_2$ of $\scri^+$, there exists a supertranslation $\mathcal{T}$ such that
\be
    \mathcal{T}^\ast (\mathcal{C}_1) = \mathcal{C}_2
\ee 
from which we deduce that $\mathcal{T}^\ast(\Delta_1) = \Delta_2$. 
In addition, the Hamiltonian commutes with the generator of a supertranslation $[\hat{H}, \hat{P}_T] = 0$ \footnote{Which should follow from the bracket $[T \p_u, \p_u] = 0$.}, following immediately from the time independence of the Hamiltonian \eqref{restricthamil}. The action of the supertranslation consists basically to reparametrize the retarded time on $\scri^+$, mapping the coordinate system locating the cross section $\mathcal{C}_1$ at $u = 0$ into an equivalent supertranslated retarded time whose origin $u' = u - T = 0$ coincides with  $\mathcal{C}_2$. Likewise, if we project the states in $\mathcal{H}_{\Delta_1}$ with compact support in $\Delta_1$ to the states in $\mathcal{H}_{\Delta_2}$ which have compact support in $\Delta_2$, then the modes $f_{\omega_0, l, m}(u)$ are now restricted to have compact support on $\Delta_2$, i.e. to $u > T$, and thus

\be
\mathcal{T}^\ast f_{\omega_0, l, m}(u) = 
\left\{
\begin{aligned}
  & \sqrt{2 \eps} Y_{lm} e^{i (\omega_0 + i \eps) u} \qquad & \text{if $u > T$}\\
  & 0 \qquad & \text{otherwise}
\end{aligned}
\right
\}
= 
\left\{
\begin{aligned}
  &\sqrt{2 \eps} Y_{lm} e^{i (\omega_0 + i \eps) u} e^{i \omega_0 T} \qquad & \text{if $u > 0$}\\
  & 0 \qquad & \text{otherwise}
\end{aligned}
\right
\}
\ee
where we still assume, of course, that $\eps$ is very small (we formally have to take the limit to $\eps \rightarrow 0$, such that $e^{- \eps T} = 1$). In this limit, $\hat{H} \lvert_{\Delta_2}$ has exactly the same form as $\hat{H} \lvert_{\Delta_1}$ and so are $\hat{\sigma} \lvert_{\Delta_2}$ and $\hat{\sigma} \lvert_{\Delta_1}$, which are related by 
\be
   \sigma \lvert_{\Delta_2} = \mathcal{T}^\ast \sigma \lvert_{\Delta_1} = U_{\mathcal{T}} \sigma \lvert_{\Delta_1} U_{\mathcal{T}}^\dag = \sigma \lvert_{\Delta_1, u' = u - T > 0}
   \label{supertranslationgibbs}
\ee
It means that if an observer turns on her detector of resolution $\eps \rightarrow 0$ at $u = 0$ and runs it during a very long time $\Delta u = \frac{1}{\eps} \rightarrow + \infty$, she will obtain the same spectrum as the observer turning on the detector at $u = T$ instead. In other words, supertranslated Hawking observers see the same thermal spectrum \footnote{As long as we are in Schwarzschild and  neglect the angular momentum potential barrier, we detail the case where we do not in section \ref{graybodysection}.}. 

\subsection*{Completing the proof}

Now, we can apply the monotonicity of relative entropy using an arbitrary state $\rho$  restricted to $\Delta_1$ and $\Delta_2$ and the reference state \eqref{gibbsatedeltai},  invoking the symmetry argument given in the previous paragraph to justify that it has the same form for $i = 1,2$. 
Therefore, we have that  
\begin{align}
    S(\rho \lvert \lvert \sigma) \lvert_{\Delta_2} &\leq S(\rho \lvert \lvert \sigma) \lvert_{ \Delta_1} \nn \\
    - S_{\Delta_2}(\rho) + \beta_H \langle \hat{H} \lvert_{\Delta_2} \rangle_{\rho} &\leq -S_{\Delta_1}(\rho) + \beta_H \langle \hat{H} \lvert_{\Delta_1} \rangle_{\rho} 
    \label{relativeentropy2}
\end{align}
where $\langle \hat{H} \lvert_{\Delta_i} \rangle_{\rho}$ is the average energy of the state $\mathcal{\rho} \lvert_{\Delta_i}$ crossing the portion $\Delta_i$ of $\scri^+$ as measured by an asymptotic observer , and $S(\rho)_{\Delta_i} = -Tr(\rho \ln{\rho})_{\Delta_i}$ is the von Neumann entropy of the state $\rho$ restricted to region $\Delta_i$. Furthermore, since we consider a free field, the radiation is extensive, and we deduce that $\langle \hat{H} \lvert_{\Delta_1} \rangle_{\rho} - \langle \hat{H} \lvert_{\Delta_2} \rangle_{\rho}$ is just the total amount of energy released on the portion $\Delta_1 \setminus \Delta_2$ of $\scri^+$ bounded by the two cross sections $\mathcal{C}_1$ and $\mathcal{C}_2$. Furthermore, through the semiclassical Einstein equations, the variation of the Bondi mass between the cuts $\mathcal{C}_1$ and $\mathcal{C}_1$ is directly related to the amount of stress energy crossing $\Delta_2 \setminus \Delta_1$ along the direction $\p_u$. Therefore, we can use the semiclassical Einstein equations and get \footnote{We can also include formally the gravitational waves in this effective stress energy tensor, by treating the gravitational perturbations through quantum fields and introducing the gravitational effective "stress energy tensor" being described by the news tensor $N_{ab}$, with $\langle T^{tot}_{ab} \rangle = \frac{1}{16 \pi G}\langle N_{ab} N^{ab} \rangle + \langle T^{matter}_{ab} \rangle n^a n^b$}
\be
    \frac{1}{8 \pi G} \Delta \mathcal{M}_i = - \int_{\mathcal{C}_1}^{\mathcal{C}_2} \langle \hat{T}_{ab} \rangle_\rho n^a n^b \eps_\scri = - (\langle \hat{H} \lvert_{\Delta_1} \rangle_{\rho} - \langle \hat{H} \lvert_{\Delta_2} \rangle_{\rho})
\ee
where  $\mathcal{M}_i$ is the geometric Bondi mass \footnote{Which is the Bondi mass up to an overall factor of $8 \pi G$. We introduced this numerical factor in order to express the Noether charges in terms of geometric quantities only, as is the area in the generalized second law.} evaluated on the cross section $\mathcal{C}_i$. Now we can write \eqref{relativeentropy2} as 
\be
    S_{\Delta_1} (\rho) - S_{\Delta_2} (\rho) + \frac{1}{8 \pi G T_H} \Delta \mathcal{M}_i \leq 0.
    \label{bondivnhorizon}
\ee
In this setting, we can see the region $\Delta_i$ of $\scri^+$ as our thermodynamic system from which we remove some energy and entropy over time by restricting the region $\Delta_1$ to the region $\Delta_2$. Now we can use another property of the von Neumann entropies. For any mutually disjoint and achronal subregions $A$ and $B$ we have that
\be \label{triangineq}
    S_{A \cup B} (\rho) \geq \lvert S_A (\rho) - S_B(\rho) \lvert \geq S_A(\rho) - S_B (\rho).
\ee
If now we choose $A = \Delta_2$ and $B = \Delta_1 \setminus \Delta_2$, then a direct application of \eqref{triangineq} leads to
\be
     \Delta \mathcal{M}_i \leq 8 \pi G T_H S_{\Delta_1 \setminus \Delta_2} (\rho)
     \label{inequalitybondidecrease}
\ee
and so, if the Bondi mass can increase, it is bounded from above by the Hawking temperature of the black hole times the von Neumann entropy of the radiation going through $\scri$ between the corners $\mathcal{C}_1$ and $\mathcal{C}_2$. It should be noticed that since the null geodesics on $\scri^+$ are complete, the achronal average null energy condition (ANEC) \footnote{which are closely related to the generalized second law \cite{Wall:2009wi}}, tells us that 
\be \label{hamiltoniandiff}
\int_{\scri^+} \langle T_{\m \n} \rangle \xi^\m n^\n\eps_{\scri^+} \geq 0
\ee
and so the total variation of the Bondi mass between $\scri^+_-$ and $\scri^+_+$ must be negative. However, it does not state that the Bondi mass cannot increase locally. On the contrary, it has already been shown that the Bondi mass can actually increase during the evaporation process, at least for two-dimensional models of evaporation \cite{Bianchi:2014qua}. 

\vspace{0.3 cm}

Now, to complete the proof of the decreasing of the $\mathcal{F}$ function, I will take a reference state $\sigma_{\mathcal{H}_{> \mathcal{B}}}$ in the future of the cross section $\mathcal{B}$ on the black hole horizon, such that
\be
    \sigma_{\Delta_i \cup \mathcal{H}_{> \mathcal{B}}} = \sigma_{\Delta_i} \otimes \sigma_{\mathcal{H}_{> \mathcal{B}}}
\ee
and let $\rho_{\mathcal{H}_{> \mathcal{B}} \cup \Delta_i}$ be a general quantum state on the region $\mathcal{H}_{> \mathcal{B}} \cup \Delta_i$ (there can be entanglement between the two regions, so that there is no particular constraint on the state $\rho_{\mathcal{H}_{> \mathcal{B}} \cup \Delta_i}$). Therefore, if we apply the inequality \eqref{relativeentropy} to the regions $A = \mathcal{H}_{> \mathcal{B}} \cup \Delta_2$  and $B = \mathcal{H}_{> \mathcal{B}} \cup \Delta_1 $ instead of $A = \Delta_2$ and $B = \Delta_1$ we find that 
\be
    S_{\Delta_1 \cup \mathcal{H}_{> \mathcal{B}}} (\rho) - S_{\Delta_2 \cup \mathcal{H}_{> \mathcal{B}}} (\rho) + \frac{1}{8 \pi G T_H} \Delta \mathcal{M}_i \leq 0.
    \label{inequalityproved}
\ee
Then, we can assume that the evolution of the quantum field is unitary $\rho \rightarrow U \rho U^\dag$, i.e  in the sense that there is no additional "coarse graining" performed by the asymptotic observer. In that case we have $S_{\Delta_2 \cup \mathcal{H}_{> \mathcal{B}}} = S_{\Sigma_2}$ and $S_{\Delta_1 \cup \mathcal{H}_{> \mathcal{B}}} = S_{\Sigma_1}$. Therefore, we find that
\be
    8 \pi G T_\mathcal{H} [S_{\Sigma_1} (\rho) - S_{\Sigma_2} (\rho)] + \Delta \mathcal{M}_{i} = \Delta \mathcal{F}_i \leq 0
    \label{lawf}
\ee
where
\be \label{freeenergy}
    \mathcal{F}_i = \mathcal{M}_i - 8 \pi G T_H S_{\Sigma_i}
\ee
is the black hole free energy \footnote{We can add the area of the corner $\mathcal{B}$ on the black hole horizon to the entropy and write $\mathcal{F}_i = \mathcal{M}_i - 8 \pi G T_H (S_{\Sigma_i} + \frac{A_\mathcal{B}}{4 G \hbar})$. This is irrelevant as we are interested in differences of this quantity along the successive spacelike cross sections which intersect the horizon at the same corner $\mathcal{B}$, and the area disappears while taking differences. As we previously claimed, the black hole area is not a relevant geometric quantity for the asymptotic observer.} . We have now completed our proof and have obtained the desired relation. An interesting point is that the classical limit of \eqref{lawf} is very similar to the the classical limit of the generalized second law. Indeed, $T_H = \frac{\hbar \kappa}{2 \pi}$ where $\kappa$ is the black hole surface gravity. Therefore, if $\hbar \rightarrow 0$, then we recover the classical Bondi mass loss formula, as we recover the classical area theorem on the event horizon if $\hbar \rightarrow 0$. However, these two theorems are not necessarily true if we take quantum effects into account. Furthermore, the analogy with the generalized second law can be pushed further, as we have now a relation between a geometric Noether charge, the geometric Bondi mass $\mathcal{M}$, and a quantity derived from quantum information, the von Neumann entropy, which are related by the factor $\hbar G$ reminiscent of quantum gravity. 

\subsection*{Comparison with the generalized second law}

In Wall's proof of the generalized second law \cite{Wall:2011hj}, the strategy is similar, but there are important differences. Assuming Lorentz symmetries and the stability condition, Wall uses the Bisongano-Wichmann theorem \cite{Bisognano:1975ih} to prove that the state of the quantum field on the future of an arbitrary cross section (say $v = 0$) on the event horizon is exactly thermal. This is basically the Unruh effect. From an operational point of view, if the observer Bob wants to collect data and measure some observables on the causal horizon, he must accelerate constantly after turning on his detector at $v = 0$. Typically, his trajectory is spanned by a boost vector field. If he does not accelerate, he will be unable to measure the observables on the horizon in the future of $v = 0$, and will just cross the horizon in a finite amount of time. On the contrary, $\scri^+$ is a horizon for the inertial observers. In addition, the Killing field, which looks like a boost close to the horizon, is a time translation far away from the black hole. Therefore, without accelerating, Alice who is far away from the black hole, will see the particles perceived by Bob which have successfully crossed the potential barrier. Therefore, in the same way as the state of the quantum field was thermal for Bob turning on his detector at $v = 0$ a long time after that the horizon settles down to a Killing horizon, it is also thermal for Alice if she turns on her detector at $u = 0$ a long time after the collapse. From this point of view, the generalized second law is a direct consequence of the Unruh effect while the monotonicity of the free energy comes from the Hawking effect \footnote{Another important similarity between the two proofs is the need for the black hole to settle down to a stationary state. In Wall's proof of the generalized second law \cite{Wall:2011hj}, this corresponds to the condition $\theta \underset{v \rightarrow + \infty}{\longrightarrow} 0$ where $\theta$ is the expansion on the horizon}. 

\vspace{0.3 cm}

Furthermore, while the generalized second law can be applied to any type of causal horizon, the law of monotonicity of free energy only makes sense in the context of a black hole. For instance, in Minkowski spacetime, the boost field is null on the Rindler horizon but the asymptotic inertial observers do not follow the orbits of the boost Killing field. They are not then immersed in a thermal bath of particles and it is therefore impossible to reproduce the proof in this context.

\section{Gray body factor}
\label{graybodysection}

In general, however, the spectrum in not exactly thermal since the fields propagate in a Schwarzschild background, inducing nontrivial gray body factors. If it is indeed true that the thermal spectrum of the vacuum fluctuations just outside the event horizon is exactly the same as the one of a black body, not all the quantum fluctuations propagate to infinity, then only a fraction of them is transmitted (the fraction $t_{\omega l m}$ for the corresponding mode) while the others are reflected by the Schwarzschild potential barrier and fall into the black hole \cite{Jacobson:2003vx}, and thus they are not detected by an asymptotic observer.  For a massless scalar field for instance,  the potential barrier of the Schwarzschild spacetime for the mode $(\omega, l, m)$ is given by
\be \label{potschw}
    V_{\omega l m} = (1 - \frac{2M}{r})(\frac{l(l + 1)}{r^2} - \frac{2M}{r^3})
\ee
in the sense that the modes $(\omega, l, m)$ satisfy a wave equation with a potential of the form \eqref{potschw}
However, if we put the black hole in a box whose walls are at the same temperature $T = T_H$ as the black hole, then the black hole ultimately reaches thermal equilibrium with the box if we disregard instabilities \footnote{Of course, instabilities occur because the energy of the black hole decreases with its temperature.}, since the absorption coefficient $T_{\omega l m}$ by the black hole of the mode $(\omega, l, m)$ is equal (in modulus) to the transmission amplitude $t_{\omega l m}$ of the outgoing modes from the past event horizon to infinity (and there is a similar relation between the reflection coefficients $R_{\omega l m}$ for the modes coming from infinity and $r_{\omega l m}$ for the modes coming from the past event horizon) \footnote{In fact, because of spherical symmetry, the reflection and transmission coefficient should not depend on the $z$ angular momentum component $m$. However, it is not the case for the Kerr black hole and therefore we find it more convenient to keep this additional index in the Schwarzschild case too.}. We can understand this statement by noticing that the potential \eqref{potschw} cancels out both at the horizon and infinity, and therefore, there is the same potential "gap" for the incoming and outgoing modes $(\omega,l, m)$. Of course, the modulus (square) of these (complex) numbers gives us scattering probabilities for the mode $(\omega, l, m)$ by the black hole geometry.  In that case, the thermal state at $\scri^+$ is therefore exactly by \eqref{gibbsstate}.

\vspace{0.3 cm}

On the contrary, in the case in which the black hole is the only source of  radiation, it is essential to take into account the gray body factors. For instance, let us consider some particle coming from infinity with energy $E = \hbar \omega$ and angular momentum $l$ propagating on a Schwarzschild background. If the angular momentum $l$ is large, then the particle will not penetrate into the black hole, because its impact parameter is too big. Therefore the transmission coefficient $T_{\omega l m}$ associated to this mode is very small in modulus compared to the reflection coefficient $R_{\omega l m}$, and so are $t_{\omega l m}$ and $r_{\omega l m}$. Likewise, we understand that for modes of very high frequency such that $\omega R_S >> 1$, the potential barrier is invisible and therefore such modes are not reflected by the geometry while if $\omega R_S << 1$, then the black hole is not able to absorb or emit such modes, and so the transmission amplitudes are very small in modulus. This analysis is strengthened by technical computations of the transmission and reflection coefficients \cite{Neitzke:2003mz,harmark2010greybody}. 
\vspace{0.3 cm}

\begin{figure}
\begin{center}
\includegraphics[scale=0.5]{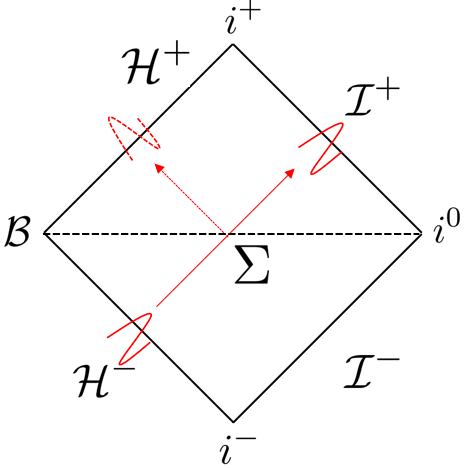}
\caption{We model here the collapse by an eternal Schwarzschild black hole (where only the region outside the horizon is depicted here) and therefore $\Sigma$ is the restriction of a Cauchy surface in the region outside the black hole. We can quantize our field using the modes $e^{i \omega v}$ and $e^{i \omega U}$, where $U = - e^{-\kappa u}$ is an affine time on the past horizon and $v$ is an affine time at $\scri^{-}$, which are defined on the black hole region and one exterior of the maximally extended Schwarzschild spacetime. The corresponding vacuum state is the Unruh state \cite{Unruh:1976db}, which is a mixed state if restricted to the black hole exterior region. On the past horizon, it can be seen as a thermal state with respect to the Bondi retarded time $u = \frac{\ln{-U}}{\kappa}$ on the past horizon. However, the modes have to propagate through the stationary background before reaching out future null infinity and the detector of the asymptotic observer. In doing so, they interact with the gravitational potential and some particles are reflected back towards the black hole and do not reach $\scri^+$. This is why the density matrix is not thermal at $\scri^+$ while it is thermal on the horizon.
}
\label{extendedschwcauchyslice}
\end{center}
\end{figure}

In order to take into account the influence of the potential properly, we should first compute the density matrix $\sigma$ of the Hawking radiation in the general case. Following Unruh \cite{Unruh:1976db}, we use a shortcut and model the Hawking radiation at late time on $\scri^+$ by considering an eternal black hole instead of a collapsing star where we have to set the following initial conditions:  we consider no incoming particles at $\scri^{-}$ and assume that the quantum fields are exactly thermal with respect to the null Killing field  on the past event horizon $\mathcal{H}^-$ (see figure \ref{extendedschwcauchyslice}) , i.e. 
\be 
        \sigma_0 = \frac{e^{- \beta_H \hat{H}}}{Tr (e^{- \beta_H \hat{H}})} =  \sum_{ \{ N_{\omega l m} \} } \prod_{\omega l m} \bigg( (1 - e^{- \beta_H \hbar \omega}) e^{- \beta_H N_{\omega l m} \hbar \omega} \ket{N_{\omega l m}} \bra{N_{\omega l m}} \bigg)
\ee
where $N_{\omega l m}$ is the number of particles with frequency $\omega$, angular momentum $l$ and vertical angular momentum component $m$. Since for a massless scalar field the modes interact with the potential barrier given by \eqref{potschw}, some particles will not be able to propagate from $\mathcal{H}^-$ to $\scri^+$ and will be reflected back into the black hole, see Figure \ref{extendedschwcauchyslice}.

Therefore, the problem consists in knowing how many particles are able to cross the potential barrier, assuming that the transmission coefficient is $t_{\omega l n}$ and the reflection coefficient is $r_{\omega l m}$. In other words, we can compute the probability $\sigma_{ N_{\omega l m} }$ of having number $N_{\omega l m}$ of particles with quantum numbers $\omega, l, m$ at $\scri^+$ by summing the probability that we had initially $N_{\omega l m}$ particles at $\mathcal{H}^-$ and that they have all crossed the potential, or that we had $N_{\omega l m} + 1$ at $\mathcal{H}^-$ and only $N_{\omega l m}$ of them have crossed the potential barrier, and so on. In other words
\begin{align} \label{realstate}
    \sigma_{N_{\omega l m}} \lvert_{\Delta_i} 
    &= (1 - e^{- \beta_H \hbar \omega}) [e^{- \beta_H \hbar \omega} \rvert t_{\omega l m} \lvert^2]^{N_{\omega l m}} \sum_{M = N_{\omega l m}}^{+ \infty} \binom{M}{N_{\omega l m}} [e^{- \beta_H \hbar \omega} (1 - \rvert t_{\omega l m} \lvert^2)]^{M - N_{\omega l m}} \nn \\
    &= (1 - e^{- \beta_H \hbar \omega}) \frac{[e^{- \beta_H \hbar \omega} \rvert t_{\omega l m} \lvert^2]^{N_{\omega l m}}}{(1 - (1 - \rvert t_{\omega l m} \lvert^2) e^{- \beta_H \hbar \omega})^{N_{\omega l n} + 1}}
\end{align}
Of course, if $\rvert t_{\omega l m} \lvert^{2} = 1$, then we recover the thermal spectrum that we considered in the previous section \ref{derivation}. However, it is clear that the correction to the thermal spectrum has consequences on the term $Tr(\rho \ln {\sigma})$ appearing in the relative entropy  \eqref{relativeentropy} since the reference state that is the image on $\mathcal{H}_{\Delta_i}$ of $\sigma_0 = \ket{0}_{-} \bra{0}_-$ is modified, and there are new terms appearing in the flux balance law obtained from the monotonicity of the relative entropy  \eqref{relativeentropy2}. Therefore, we must give a physical interpretation to these additional terms in our formulas. 

 \subsection*{Chemical potential}

There exists a way to take into account the presence of the gray body factors that is intuitive and which pushes further the interpretation of the flux balance law at $\scri^+$ as the spontaneous evolution of an ordinary thermodynamic potential. We define a \textit{chemical potential} $\m_{\omega l m}$ associated to chemical species $(\omega, l, m)$ [i.e. modes defined by the quantum numbers $(\omega, l, m)$] such that 
\be \label{chemicalpot}
    \m_{\omega l m} :=  T_H \ln{\frac{\rvert t_{\omega l m} \lvert^2}{1 - (1 - \rvert t_{\omega l m} \lvert^2) e^{-\beta_H \hbar \omega}}}
\ee
It is not the first time that a definition for a chemical potential has been proposed for black hole thermodynamics, see, for instance, \cite{Dolan:2014cja, Maity:2015ida, Kubiznak:2016qmn, visser2022holographic} in the context of AdS/CFT. However, to the best of my knowledge, this is the first time that the definition \eqref{chemicalpot} is proposed.

\vspace{0.3 cm}

There are several remarks that need to be made at this stage. First, the chemical potential is negative, as it is the case for a classical gas or a gas of bosons. Second, it vanishes if the modulus of the transmission coefficient is equal to $1$, i.e. if the distribution of the mode is exactly thermal. From \eqref{realstate}, we can now write the reference state $\sigma$ as 
\begin{align} \label{gibbstatechemical}
        \sigma_{\{ N_{\omega l m} \} } \lvert_{\Delta_i} &= \prod_{\omega l m} \frac{(1 - e^{- \beta_H \hbar \omega})}{(1 - (1 - \rvert t_{\omega l m} \lvert^2) e^{- \beta_H \hbar \omega})} e^{- \beta_H (E_{\omega l m} - \m_{\omega l m} N_{\omega l m})} = \prod_{\omega l m} \mathcal{N}_{\omega l m} e^{- \beta_H (E_{\omega l m} - \m_{\omega l m} N_{\omega l m})} \nn \\
        &= \mathcal{N} e^{- \beta(E - \sum_{\omega l m} \m_{\omega l m} N_{\omega l m})} 
\end{align}
where $E_{\omega l m} = N_{\omega l m} \hbar \omega$ is the energy of $N_{\omega l m}$ particles in the mode $(\omega, l, m)$, $E = \sum_{\omega l m} E_{\omega l m}$ is the total energy, and $\mathcal{N}_{\omega l m}$ and $\mathcal{N}$ are normalization factors. Therefore, the distribution \eqref{gibbstatechemical} follows the grand canonical ensemble distribution and not the one of the canonical ensemble anymore. The potential barrier acts exactly as the source of a chemical potential through the induced nontrivial gray body factors since the number of Hawking particles is not conserved between the vicinity of the horizon (where they are created) and future null infinity. Alternatively, the chemical activity 
\be
        a_{\omega l m} := e^{\beta_H \m_{\omega l m}}= \frac{\rvert t_{\omega l m} \lvert^2}{1 - (1 - \rvert t_{\omega l m} \lvert^2) e^{-\beta_H \hbar \omega}}
\ee
is equal to $1$ if the modes $(\omega, l, m)$ do not interact with the gravitational potential, which is the case in the standard state $M \rightarrow + \infty$.

\vspace{0.3 cm}

More precisely, we can proceed as we did in order to compute the correction from the gray body factors to the spectrum and replace the picture of the collapsed black hole with the picture of an eternal black hole where a thermal distribution is introduced on the past horizon horizon while there are no incoming particles at $\scri^-$. In this picture, some particles are produced on the past horizon (i.e in the vicinity of the event horizon for a real collapse scenario) but their number is not conserved, since they interact with the gravitational potential and some of them disappear, being absorbed by the black hole before reaching out $\scri^+$.  This nonconservation of the number of particle is responsible for the appearance of a chemical potential in the formulas. Following the same lines of the derivation in Section \ref{derivation} \footnote{with the exception that now $\mathcal{C}_1$ and $\mathcal{C}_2$ must be related by an arbitrary time translation and not by an arbitrary supertranslation. This is because the gray body factors depend on the angular momentum while an arbitrary supertranslation changes the foliation of $\scri^+$, and so shifts the angular momentum for the same reason as a usual translation acts non trivially on the angular momentum charge by shifting the origin with respect to which the angular momentum is evaluated. Indeed, the Lorentz group is neither an ideal of the BMS group nor an ideal of the Poincare group, and so there is no preferred choice of Lorentz subalgebra, which means that to each choice of origin or foliation of $\scri^+$ there is a different notion of angular momentum. In the present case, we want to evaluate the angular momentum of the wave packets with respect to the black hole "origin" $r = 0$, and therefore we do not allow supertranslated shifts of cross sections at $\scri^+$.}, we can prove that between two arbitrary cross sections at late time on $\scri^+$ related by a global time translation

\be \label{newfreeeenrgyineq}
\Delta \mathcal{F}_i = \Delta \big( \mathcal{M}_i - 8 \pi G T_H S_{\Sigma_i} - 8 \pi G \sum_{\omega, l, m} \m_{\omega, l, m} \langle \hat{N}_{\omega l m} \rangle_{\rho} \big)  \leq 0
\ee
and where the free energy $\mathcal{F}$ is given by

\be \label{newfreeenergy}
    \mathcal{F}_i = \mathcal{M}_i - 8 \pi G T_H S_{\Sigma_i} - 8 \pi G \sum_{\omega, l, m} \m_{\omega, l, m} \langle \hat{N}_{\omega l m} \rangle_{\rho, i}
 \ee
expressed with the same physical quantities as in \eqref{freeenergy} but with an additional conjugate pair involving the number of particles and the associated chemical potential \eqref{chemicalpot} for each mode. However it is important to notice that the formula \eqref{newfreeenergy} is not well defined, even if we do not consider the divergences due to the presence of entanglement entropy, because of the term $ \langle \hat{N}_{\omega l m} \rangle_{\rho, i}$. Indeed, $\langle \hat{N}_{\omega l m} \rangle_{\rho, i}$ is supposed to be the number of particles in the mode $(\omega, l, m)$ on the slice $\Sigma_i$, but the notion of particle is not even defined if we are not at $\scri^+$ since spacetime is not stationary. However, if we consider the system to be the region $\Delta_i$ of $\scri^+$ instead of $\Sigma_i$, then the number of particle operators are well defined and \eqref{newfreeenergy} makes sense (of course, we also have to replace $S_{\Sigma_i}$ by $S_{\Delta_i}$ and regularize the eventual divergences). 

\vspace{0.3 cm}

Nevertheless, if we consider that the slices $\Sigma_i$ intersect the event horizon at a late time cross section $\mathcal{B}$ such that the black hole has already settled down to the background Schwarzschild solution, then we can assume that the number of particles is conserved between the slice $\Sigma_i$ and the regions $\Delta_i$ and $\mathcal{H}_{> \mathcal{B}, i}$, the latter being the portion of the event horizon in the future of the region $\mathcal{B}$. In this case, $\langle \hat{N}_{\omega l m} \rangle_{\rho, i}$ can be roughly interpreted as the number of outgoing particles on $\Sigma_i$. Of course, not all of them reach future null infinity because of the gravitational potential and this explains the presence of the chemical potential. The picture is more reliable if we look at quantum fields propagating in an eternal black hole spacetime and assume that there is no incoming radiation from $\scri^-$ so that the quantum state $\rho$ of the field at $\scri^+$ is obtained after that the modes have propagated from the past horizon to future null infinity. Since the background spacetime is stationary all along, the number of particles is conserved between $\mathcal{H}^{-}$ and $\scri^+$ except for the ones falling into the black holes because of the backscattering. 

\subsection*{Equilibrium configuration : Hawking radiation}

We can notice that this is necessary to take into account the gray body factors in order to treat consistently Hawking radiation. Indeed, we can apply the formula \eqref{newfreeeenrgyineq} to the state $\rho = \sigma$ on an arbitrary late time portion of $\scri^+$. In this case it is obvious that the inequality is saturated since the relative entropy vanishes and we have \footnote{Since the radiation is thermal the particles are uncorrelated and therefore we state that $S_{\Sigma_2}(\sigma) - S_{\Sigma_1}(\sigma) = S_{\Delta_1 \setminus \Delta_1} (\sigma)$.} 
\begin{align} \label{freeeenrgyhw}
    \Delta (\mathcal{M}_i - 8 \pi G T_H S_{\Delta_i}) &= \Delta \big(8 \pi G \sum_{\omega l m} \m_{\omega l m}  \langle N_{\omega l m} \rangle_{\Delta_i} \big) = - 8 \pi G \sum_{\omega l m} \langle N_{\omega l m} \rangle_{\Delta_1 \setminus \Delta_2} T_H \ln{\frac{\rvert t_{\omega l m} \lvert^2}{1 - (1 - \rvert t_{\omega l m} \lvert^2) e^{-\beta_H \hbar \omega}}} \nn \\
        &= - 8 \pi G T_H\int_{\mathcal{C}_1}^{\mathcal{C}_2} \sum_{\omega l m} \frac{\rvert t_{\omega l m} \lvert^2}{e^{\beta_H \hbar \omega} - 1}  \ln{\frac{\rvert t_{\omega l m} \lvert^2}{1 - (1 - \rvert t_{\omega l m} \lvert^2) e^{-\beta_H \hbar \omega}}} du \geq 0
\end{align}
so that the free energy variation without taking into account the chemical potentials and the variation of the particle numbers is positive. However, this is an expected result. Indeed, we have that $\Delta \mathcal{M}_i = T_H \frac{\Delta \mathcal{A}}{4 G \hbar}\leq 0$ where $\Delta \mathcal{A}$ is the area variation of the event horizon \footnote{Notice here that the cross sections of the event horizon between which the area variation $\Delta A$ is evaluated are left unspecified, since these cross sections are not causally related to $\scri^+$ and so there is no way of saying that the mass variation $\Delta M$ of Bondi mass between two cross sections $\mathcal{C}_1$ and $\mathcal{C}_2$ at null infinity is related to the area variation between two cross sections $\mathcal{S}_1$ and $\mathcal{S}_2$ at $\scri^+$.} that we can deduce from the outgoing energy flux at null infinity by energy conservation and therefore we can write  \eqref{freeeenrgyhw} as
\be
    \Delta (\frac{\mathcal{A}}{4 G \hbar} + S_{\Delta_1 \setminus \Delta_2}) = -\int_{\mathcal{C}_1}^{\mathcal{C}_2} \sum_{\omega l m} \frac{\rvert t_{\omega l m} \lvert^2}{e^{\beta_H \hbar \omega} - 1}  \ln{\frac{\rvert t_{\omega l m} \lvert^2}{1 - (1 - \rvert t_{\omega l m} \lvert^2) e^{-\beta_H \hbar \omega}}} du  \geq 0.
\ee
We recover the usual result that the entropy of Hawking radiation is larger than the area reduction coming from the negative energy flux of matter into the black hole. We can notice that the entropy creation term vanishes if $\lvert t_{\omega l m} \rvert^2$ vanishes or is equal to $1$. In the first case, this is because no radiation reaches out null infinity and in the second case this is because the entropy of the radiation emitted by a black body is directly related to its energy, as it is the case for a perfect black body \footnote{We can use the positivity of the relative entropy $S(\rho \lvert \lvert \sigma)$ to show that the total variation of free energy on $\scri^+$ is negative (or alternatively that the free energy defined on $\scri^+$ is positive). In this case, $\rho = \sigma$, the relative entropy vanishes and we have that $
(\mathcal{M}_{+ \infty} - \mathcal{M}_{\mathcal{C}_i}) - T_H S_{\Delta_i} (\sigma) = - \ln{Z} - \sum \m_{\omega l m} \langle N_{\omega l m} \rangle_\rho \geq 0$. We can then set that the total mass variation comes from Hawking radiation and causes the area to shrink, with $\Delta \mathcal{M} = T_H
 \frac{\Delta \mathcal{A}}{4 G \hbar}$. Therefore, we also recover here that the generalized second law increases  when we consider the total flux at $\scri^+$. However, this total flux involves the final mass $\mathcal{M}_{+ \infty}$, which is needed to be assumed and the constant $- \ln{Z}$.  Looking at differences between arbitrary cross sections allows one to get rid of these constants. }. 
In addition, since $d \mathcal{F}_i = 0$, corresponding to no spontaneous evolution, we can write \eqref{freeeenrgyhw} as
 \be
        d \mathcal{M}_i  = 8 \pi G T_H d S_{\Delta_i} + 8\pi G \sum_{\omega l m} \m_{\omega l m} d \langle N_{\omega l m} \rangle_{\Delta_i}
 \ee
 and in this case we can define the chemical potential by
 \be
        8 \pi G \m_{\omega l m} = \frac{\p \mathcal{M}}{\p \langle N_{\omega l m} \rangle} \Big\vert_{S, \langle N_{\omega' l' m'} \rangle \neq \langle N_{\omega l m} \rangle}
 \ee
as we usually do for equilibrium configurations of ordinary chemical systems.

\subsection*{Classical limit}
\label{classicallimit}

The nonvanishing gray body factors also play a role in the classical limit of the formula \eqref{newfreeeenrgyineq}. Indeed, we can write \eqref{gibbstatechemical} as 
\be
    \sigma_{\{ N_{\omega l m} \} } \lvert_{\Delta_i} = \prod_{\omega l m} \mathcal{N}_{\omega l m} e^{- \beta_H E'_{\omega l m}}
\ee
where $E'_{\omega l m} = N_{\omega l m} \hbar \omega' \geq N_{\omega l m} \hbar \omega = E_{\omega l m}$ in which 
\be
    \omega' = \omega - \hbar^{-1} \m_{\omega l m} = \omega - \frac{\kappa}{2 \pi} \ln{a_{\omega l m}} \geq \omega \geq 0.
\ee
In the classical limit we have $\hbar \rightarrow 0$ and $\m_{\omega l m} \rightarrow 0$ since the chemical potential is proportional to the Hawking temperature $T_H$ and therefore proportional to $\hbar$. We also have
\be
\sum_{\omega - \eps \leq \omega \leq \omega + \eps, l m}\langle N_{\omega l m} \rangle_\rho \rightarrow + \infty
\ee
in order to have a finite energy density $\hat{E}_{lm}(\omega) \vert_\rho = \hbar \omega \langle N_{\omega l m} \rangle_\rho$ around the frequency $\omega$ with uncertainty $\eps << \omega$. If we assume in the classical limit that the von Neumann entropy variation of the quantum fields is of order $O(1)$ \footnote{Do not forget that we took the Boltzmann constant to be $k_B = 1$ in our analysis}, then the classical limit of \eqref{newfreeeenrgyineq} is 
\be \label{freeenergyvarclass}
    \Delta \mathcal{F} \vert_\rho \approx \langle \hat{H}' \vert_{\Delta_1}\rangle_\rho - \langle \hat{H}' \vert_{\Delta_2}  \rangle_\rho
\ee
where $\hat{H}' \lvert_{\Delta_i}$ is an operator diagonal in the same basis as $\hat{H} \lvert_{\Delta_i}$ but with eigenvalues shifted from $E_{\omega l m}$ to $E_{\omega l m}'$. Of course, \eqref{freeenergyvarclass} is not equal to the variation of the Bondi mass $\Delta \mathcal{M} \vert_\rho$ in general. However we can consider another state $\rho'$ for the outcoming radiation such that $\rho_{\{ N_{\omega l m} \}} = \rho_{\{ N_{\omega' l m} \}}'$ for any mode $(\omega, l, m)$ so that $\langle \hat{H}' \vert_{\Delta_i} \rangle_\rho  = \langle \hat{H} \vert_{\Delta_i} \rangle_{\rho'}$, and therefore, 
\be
    \Delta \mathcal{F} \vert_\rho = \Delta \mathcal{M} \vert_{\rho'} \leq 0
\ee
in the classical limit. Of course, if we consider some radiation peaked around the frequency $\omega$ such that $\omega >> \hbar^{-1} \m_{\omega l m}$, then $\omega' = \omega$ and $ \Delta \mathcal{F} \vert_\rho$ reduces to $\Delta \mathcal{M} \vert_\rho$. This is expected to be true if $\omega R_S >> \ln{\lvert t_{\omega l m} \rvert^2}$. 

\subsection*{Effective temperature}

Another way of thinking would be to follow Page \cite{Page:2004xp} and associate a different temperature to each mode $T_{\omega l m}$ such that $T_{\omega l m} = T_H$ if and only if $\lvert t_{\omega l m} \rvert = 1$. This effective temperature is, of course, related to the chemical potential of the mode $(\omega l m)$ by the relation
\be
    T_{\omega l m} = \frac{T_H}{ (1 - \frac{\m_{\omega l m}}{\hbar \omega})} \leq T_H
\ee
Therefore, we can write the free energy spontaneous evolution law as
\be \label{multipletemp}
    \Delta F = \sum_{\omega l m} \frac{\Delta \langle E_{\omega l m} \rangle_\rho}{T_{\omega l m}} - \Delta S_{\Sigma_i} (\rho) \leq 0
\ee
and thus we should interpret the black hole surrounding as an ensemble of reservoirs with a different temperature $T_{\omega l m}$ for each mode. The drawback of this picture is that the geometric Bondi mass does not appear in the formula. In addition, the temperature of the black hole is uniquely defined as $T_H$, since the gravitational system is in thermal equilibrium with a gas at this temperature, and so it seems less natural to introduce these effective temperatures $T_{\omega l m}$. Thus, we will stick with the chemical potentials and the formula \eqref{newfreeeenrgyineq} instead of the equivalent formula \eqref{multipletemp} because the former seems to be more physically reliable. 

\section{Kerr black hole}
\label{kerrsection}

In the case of the Kerr black hole, the discussion is very similar to the one we had for the Schwarzschild black hole. The black hole horizon now has a an "angular velocity" $\Omega_H$ with respect to the asymptotic observers. Hence, we work now with fixed black hole temperature $T_H = \frac{\hbar \kappa}{2 \pi}$ and angular velocity $\Omega_H$. Of course, the situation is also very similar to the standard thermodynamic case where both the temperature and the chemical potential are held fixed. Indeed, Hawking's computation \cite{hawking1975particle} shows that at very late time $u \rightarrow + \infty$ the asymptotic observer sees an average number of particles in the mode $(\omega, l, m)$ equal to
\be
    \langle N_{\omega, l, m} \rangle = (1 - \frac{m \Omega_\mathcal{H}}{\omega}) \lvert t_{\omega, l, m} \rvert^2 \frac{1}{e^\frac{2 \pi (\omega - m \Omega_H)}{\kappa} - 1} =  \frac{\Gamma_{\omega, l, m}}{e^\frac{2 \pi (\omega - m \Omega_H)}{\kappa} - 1}
\ee
which, if we put aside the modified gray body factor 

\be
\Gamma_{\omega, l, m} = (1 - \frac{m \Omega_\mathcal{H}}{\omega}) \lvert t_{\omega, l, m} \rvert^2 = 1 - \lvert r_{\omega l m} \rvert^2
\label{reflectioncoeff}
\ee
is also the average number of particles in the mode $(\omega, l, m)$ in the grand canonical ensemble with chemical potentials $m \Omega_H$. Of course, the gray body factor \eqref{reflectioncoeff} is negative for superradiant modes $\omega < m \Omega_H$. However, as for the nonrotating black hole, we need more than the average number of particles in each mode and we have to compute the quantum state of the field at very late time on $\scri^+$. The reference state can be obtained \cite{hawking1975particle ,Wald:1975kc, Page:2004xp} from \eqref{realstate}  by shifting $\omega$ into $\omega - m \Omega_H$ and $\lvert t_{\omega l m} \rvert^2$ into $\Gamma_{\omega l m}$

\be \label{realstatekerr}
\sigma_{\{ N_{\omega l m} \}} \lvert_{\Delta i} = \prod_{\omega l m} (1 - e^{- \beta_H \hbar (\omega - m \Omega_H )}) \frac{[e^{- \beta_H \hbar (\omega - m \Omega_H )} \Gamma_{\omega, l, m}]^{N_{\omega l m}}}{(1 - (1 - \Gamma_{\omega, l, m}) e^{- \beta_H \hbar (\omega - m \Omega_H )})^{N_{\omega l n} + 1}}
\ee
The reference state \eqref{realstatekerr} is the one to consider for the most generalized analysis. However, as for Schwarzschild, we can proceed in two steps and neglect the gray body factors in the first instance.

\subsection*{Neglecting the gray body factors}

We consider the modes $(\omega, l, m)$ in the range for which $\Gamma_{\omega, l, m} \sim 1$, a condition satisfied for sufficiently high frequencies and low angular momentum, because then the wave packet is blind to the potential barrier induced by the Kerr geometry. In this range of frequency and angular momentum, the reference state \eqref{realstatekerr} simplifies to 

\be
    \sigma = \frac{e^{- \beta_H (\hat{H} - \Omega_H \hat{J})}}{Tr (e^{- \beta_H (\hat{H} - \Omega_H \hat{J})})}
    \label{thermalgrandcano}
\ee
and the Kerr black hole has a temperature $T_H$, even if the probability distribution is the one of the grand canonical ensemble and not the one of the canonical ensemble. In some sense, it is similar to what we had in Section \ref{graybodysection}, but the physical interpretation is different, since the asymptotic angular momentum is not identified to an average number of particles. Therefore, we just state that the probability distribution is shifted compared the one of the canonical ensemble because there is an additional nonvanishing intensive parameter with a mechanical origin, the angular velocity $\Omega_H$. In this case we can follow the same lines as we did for the Schwarzschild black hole \footnote{With the caveat that the vertical angular momentum generator $\hat{J}_z$ and the generator of an arbitrary supertranslation $\mathcal{P}_\mathcal{T}$ do not commute, and so as in the case of the Schwarzschild black hole with angular momentum dependent gray body factors, we consider only global time translations here, so that the cross sections $\mathcal{C}_1$ and $\mathcal{C}_2$ belong to the same $u$ foliation.}
as long as we consider a quantum state of the field $\rho$ with characteristic frequency and angular momentum belonging to the range $\Gamma_{\omega l m} \sim 1$. We find that 
\be
\Delta (\mathcal{M}_i - \Omega_H \mathcal{J}_i - 8 \pi G T_H S_{\Sigma_i}) \leq 0
\label{grand potentiel inequality}
\ee
where $\mathcal{J}_i$ is the geometric Dray-Streubel angular momentum \footnote{The flux balance laws at $\scri^+$ allow us to write that the variation of the geometric Dray-Streubel charges between two cross sections $\mathcal{C}_1$ and $\mathcal{C}_2$ is 
\be
        \frac{1}{8 \pi G} \Delta \mathcal{J}_i = \frac{1}{8 \pi G} \int_{\mathcal{C}_1}^{\mathcal{C}_2} N^{ab} \mathcal{L}_{\p_\phi} \sigma_{ab} \eps_\scri - T_{ab} \p_\phi^a n^b \eps_\scri
\ee
where $\sigma_{ab}$ is the asymptotic shear and $\p_\phi$
is the axisymmetric background Killing field. Here we can replace $T_{ab}$ by a quantum operator and the piece $N^{ab} \mathcal{L}_{\p_\phi} \sigma_{ab}$ can be interpreted as the angular momentum flux of the gravitons.} evaluated on a cross section $\mathcal{C}_i$ of $\scri^+$. Therefore, the appropriate thermodynamic potential (which we still call free energy by convention) is $\mathcal{F}_i = \mathcal{M}_i - \Omega_H \mathcal{J}_i - 8 \pi G T_H S_{\Sigma_i}$ and decreases over time. The classical limit $\hbar \rightarrow 0$ gives that $\Delta (\mathcal{M}_i - \Omega_H \mathcal{J}_i) \leq 0$ which might seem a bit surprising, since we know that if the Bondi mass decreases in the classical limit, then the angular momentum can take arbitrary (positive or negative) values. However, we have to keep in mind that the formula \eqref{grand potentiel inequality} is valid as long as the reference state $\sigma$ is thermal, i.e. that $\sigma$ is given by \eqref{thermalgrandcano} which is true for the modes satisfying $\Gamma_{\omega, l, m} \sim 1$. In this regime, we expect $\Delta (\mathcal{M}_i - \Omega_H \mathcal{J}_i) \leq 0$ to be true classically. Indeed, in a Kerr background, the transmission and reflection amplitudes of a wave packet are given by \eqref{reflectioncoeff} and thus $\Gamma_{\omega l m} \sim 1$ only for large frequencies $\omega >> m \Omega_H $.

\subsection*{Including the gray body factors}

Now, if we want to make our results more precise we have to compare an arbitrary state $\rho$ with the reference state $\sigma$ \eqref{realstatekerr} that we obtain from evolving the in-vacuum state on $\scri^-$ to late time on $\scri^+$. The monotonicity of the relative entropy gives us that
    \be \label{more general potential}
\Delta \big( \mathcal{M}_i - \Omega_H \mathcal{J}_i - 8 \pi G T_H S_{\Sigma_i} - 8 \pi G\sum_{\omega, l, m} \m_{\omega l m} \langle \hat{N}_{\omega l m} \rangle_{\rho} \big) \leq 0
\ee
using the same definition of the chemical potential as in \eqref{chemicalpot} by shifting $\omega$ into $\omega - m \Omega_H$ and $\lvert t_{\omega l m} \rvert^2$ into $\Gamma_{\omega l m}$ (the chemical potential is still negative). It can be noticed that \eqref{realstatekerr} can be written as 

\be \label{realstatekerr1}
\sigma_{\{ N_{\omega l m} \} } \lvert_{\Delta_i} = \prod_{\omega l m} \mathcal{N}_{\omega l m}  e^{- \beta_H N_{\omega l m} \hbar \omega'}
\ee
where the $\mathcal{N}_{\omega l m}$ are the normalization factors for each mode and 

\be \label{omprimede}
    \omega' = \omega - m \Omega_H - \hbar^{-1} \m_{\omega l m} \geq 0
\ee
which is indeed positive even for the superradiant modes $\omega - m \Omega_H \leq 0$ as it can be checked after a quick analysis. Therefore, the classical limit of \eqref{more general potential} is not different from the one of the Schwarzschild black hole studied in Section \ref{classicallimit} with the appropriate substitution \eqref{omprimede} for $\omega'$. 
Now we can also see what happens if we take the limit $\kappa \rightarrow 0$ for extremal Kerr black holes. In order to do this, we can directly look at \eqref{realstatekerr} and take the limit. If $\omega - m \Omega_H > 0$, then we have that
\be \label{normalmodeslimit}
\sigma_{\{ N_{\omega l m} \} } \vert_{\Delta_i} \sim \prod_{\omega l m}  [e^{- \beta_H \hbar (\omega - m \Omega_H )} \Gamma_{\omega, l, m}]^{N_{\omega l m}}
\ee
and the discussion is not very different from the Schwarzschild case discussed in Section \ref{derivation}, except that we have to change $\omega$ into $\omega' = \omega - \Omega_H m > 0$ and therefore we obtain \eqref{grand potentiel inequality} since the chemical potentials are also proportional to $\kappa$ \footnote{Of course, $\kappa \rightarrow 0$ is equivalent to the vanishing temperature limit $T_H \rightarrow 0$ if we do not tune the constant $\hbar$ such that $\hbar \kappa \rightarrow O(1)$ even though $\kappa \rightarrow 0$.}. However, we need to be a little more careful when we treat the superradiant modes $\omega - m \Omega_H < 0$. Indeed, the limit $\kappa \rightarrow 0$ gives
\be
    \sigma_{ \{ N_{\omega l m} \} } \vert_{\Delta_i} \sim \prod_{\omega l m} \frac{1}{1 - \Gamma_{\omega l m}} \big( \frac{\Gamma_{\omega l m}}{\Gamma_{\omega l m} - 1} \big)^{N_{\omega l m}} 
\ee
which is totally different from \eqref{normalmodeslimit}. In particular, the difference of relative entropy between two cross sections $\mathcal{C}_1$ and $\mathcal{C}_2$ of $\scri^+$ reduces to
\be \label{superradin}
\Delta [- \sum_{\omega l m} \langle N_{\omega l m} \rangle_\rho \ln{\frac{\Gamma_{\omega l m}}{\Gamma_{\omega l m} - 1}} - S_{\Sigma}(\rho)] \leq 0
\ee
where, of course, $\ln{\frac{\Gamma_{\omega l m}}{\Gamma_{\omega l m} - 1}} \leq 0$. 

If we consider now a charged black hole, then similar computations lead to an additional term $- \Phi_H \Delta Q_i$ to \eqref{more general potential} where $\Delta Q_i$ is the amount of electric charge flowing on the portion of $\scri^+$ of interest and $\Phi_H$ is the black hole electrostatic potential. Of course, the black hole temperature and gray body factors (and so the chemical potential) must be changed appropriately. However, there does not exist a massless charged field and therefore taking into account this term in the spontaneous evolution law of $\scri^+$ is pointless. 

\section{Outlooks}

In this paper, we have studied the nonequilibrium thermodynamics of a black hole from the perspective of a far away observer. This observer can measure the black mass and angular momentum variations, but is insensitive to the area variations. In general, we have shown that there exists a physical quantity analogous to a thermodynamic potential that can be constructed out of the Noether charges at null infinity, like the mass and angular momentum, the number of particles of each mode, and intensive thermodynamic quantities characterizing the black hole background, as the temperature, horizon velocity and chemical potential of the modes. Therefore, the main result of this analysis is the inequality
 \be \label{moregeneralpotential}
\Delta  \big( \mathcal{M} - \Omega_H \mathcal{J} - 8 \pi G T_H S_{\Sigma_i} - 8 \pi G\sum_{\omega, l, m} \m_{\omega l m}\langle \hat{N}_{\omega l m} \rangle_{\rho} \big) \leq 0
\ee
holds between arbitrary cross sections at late time of $\scri^+$ in a background spacetime of a black hole collapse.
Therefore, unlike the generalized second law which is obtained by looking at the dynamics on the horizon, the spontaneous decreasing of the free energy is obtained from the dynamics at late time on $\scri^+$.  This formula is totally consistent with the fact that a black hole behaves exactly as an open system with some temperature $T_H$ subject to a chemical reaction. 

\vspace{0.3 cm}

A generalization of this study is the application of the generalized second law 
to the case where we can simply consider two spacelike hypersurfaces located in the region outside the black hole, one of which is in the future of the other. Thus we have to look at the combined dynamics of both the horizon $\mathcal{H}$ and future null infinity $\scri^+$. Instead of considering two hypersurfaces cutting the horizon on two arbitrary cross sections and having a common endpoint at infinity as for the generalized second law or two spacelike hypersurfaces cutting null infinity at two arbitrary cross sections and having a common endpoint on the horizon as for \eqref{moregeneralpotential}, we could look at two spacelike slices $\Sigma_1$ and $\Sigma_2$ each intersecting a cross section on $\mathcal{H}$ and $\scri^+$, but such that $\Sigma_2$ is in the future of $\Sigma_1$. Since in this picture the slice $\Sigma_i$ intersects the black hole horizon at an arbitrary cut, the area of the given cross section on the event horizon should be added to the ingredients already appearing in \eqref{moregeneralpotential} to take into account the horizon dynamics. We leave this question for future work.

\vspace{0.3 cm}

Another interesting question is to understand what might happen to the free energy of the black hole when we no longer simply consider arbitrary perturbations around the equilibrium state, but when we take into account the evaporation of the black hole and therefore the backreaction. It should first be noted that in this case, according to what we recalled in Section \ref{basictherm}, the free energy is no longer a good thermodynamic potential even in ordinary thermodynamics, and that it can increase instead of decrease if the temperature of the reservoir decreases. However, during evaporation, the temperature of the black hole increases and in this case the free energy continues to decrease. Nevertheless, even if we restrict ourselves to the spherically symmetric case, we must also take into account the chemical potentials that depend on the gray body factors, which are themselves functions of the background structure. This case requires a more detailed calculation, which it might be interesting to carry out.
\section*{Acknowledgements}

I thank Luca Ciambelli, Leonard Ferdinand, Sophie Mutzel, Alejandro Perez, Simone Speziale, Salvatore Ribisi, Carlo Rovelli, Bob Wald, Sami Viollet, Victor Zhang and especially Aron Wall for useful discussions.

\bibliographystyle{unsrt}
\bibliography{bibliographe.bib}

\end{document}